\documentclass[useAMS,usenatbib,onecolumn,usegraphicx]{mn2e}

\newcommand{\lsim}{\, \, \raisebox{-0.8ex}{$\stackrel{\textstyle <}{\sim}$ }}

\newcommand{\beq}{\begin{equation}}
\newcommand{\eeq}{\end{equation}}
\newcommand{\beqar}{\begin{eqnarray}}
\newcommand{\eeqar}{\end{eqnarray}}

\title[Neutron star crustal structures]  
{Probing crustal structures from neutron star compactness}
\author[H. Sotani, K. Iida, \& K. Oyamatsu]
{Hajime Sotani$^1$ \thanks{E-mail:sotani@yukawa.kyoto-u.ac.jp},
Kei Iida$^2$, and
Kazuhiro Oyamatsu$^3$
\\
$^1$Division of Theoretical Astronomy, National Astronomical Observatory of Japan, 2-21-1 Osawa, Mitaka, Tokyo 181-8588, Japan\\
$^2$Department of Mathematics and Physics, Kochi University, 2-5-1 Akebono-cho, Kochi 780-8520, Japan\\
$^3$Department of Human Informatics, Aichi Shukutoku University, 2-9 Katahira, Nagakute, Aichi 480-1197, Japan}

\begin{document}
\maketitle
\label{firstpage}

\begin{abstract}
With various sets of the parameters that characterize the
equation of state (EOS) of nuclear matter, we systematically examine the 
thickness of a neutron star crust and of the pasta phases contained 
therein.  Then, with respect to the thickness of the phase of
spherical nuclei, the thickness of the cylindrical phase, and the crust 
thickness, we successfully derive fitting formulas that 
express the ratio of each thickness to the star's radius as a 
function of the star's compactness, the incompressibility of symmetric 
nuclear matter, and the density dependence of the symmetry energy. 
In particular, we find that the thickness of the phase of spherical nuclei
has such a strong dependence on the stellar compactness as the 
crust thickness, but both of them show a much weaker dependence 
on the EOS parameters. Thus, via
determination of the compactness, the thickness of the phase of
spherical nuclei as well as the crust thickness can be constrained 
reasonably, even if the EOS parameters remain to be 
well-determined.
\end{abstract}


\section{Introduction}
\label{sec:I}

Neutron stars help to probe the physics in extreme conditions  
mainly because the star is so compact that the density inside the star can 
be significantly beyond normal nuclear density \citep{NS}. 
Moreover, the surface magnetic field can be as high as 
$\sim 10^{15}$ G \citep{K1998,H1999}, while the rotation period can be 
as short as $\sim$msec \citep{PG2016}.  Thus, observations of neutron 
star phenomena associated with such compactness, high magnetic fields, and/or
rapid rotation could leave an imprint of the properties of matter
under such extreme conditions.  However, the neutron star structure has 
yet to be fixed, because the equation of state (EOS) of matter in the
star is still uncertain especially for  a high density region.  Even 
so, a conceptual picture of the neutron star structure is 
theoretically established.

Just below the star's surface lies an ocean composed of iron, under 
which matter forms a lattice structure due to the Coulomb interaction.  This 
region is called a crust, where the matter behaves as a solid (or 
as a liquid crystal). The region below the crust corresponds to 
a core, where the matter becomes uniform and behaves as a fluid.  The 
density at the base of the crust is expected to lie between 
$\sim(1/3$--$1)$ times normal nuclear density, depending on the EOS 
of nuclear matter \citep{OI2007}.  This EOS is often characterized by several 
parameters that determine the Taylor expansion with respect to the nucleon 
density and neutron excess around the saturation point of symmetric nuclear 
matter \citep{L1981}, which in turn can be constrained from terrestrial 
nuclear experiments \citep{OI2003,Tsang2012}.  One of the key parameters 
that control the properties of matter in the crust thickness is known 
to be the slope parameter $L$ 
of the symmetry energy \citep{OI2007}, which has yet to be fixed 
\citep{Li2017}.  This means that one may be able to extract the value of 
$L$ from astronomical observations.  In fact, after the discoveries of 
quasi-periodic oscillations in the soft-gamma repeaters \citep{WS2006}, 
attempts to constrain $L$ have been done by identifying the observed 
frequencies as the crustal torsional oscillations 
\citep{SW2009,GNJL2011,SNIO2012,SNIO2013a,SNIO2013b,S2014,S2016,SIO2016}.

Additionally, the possible presence of non-spherical (pasta) nuclei
in the deepest region of the crust of cold neutron stars has been theoretically considered
since \cite{LRP1993,O1993} (see also \cite{PR1995} for early studies on pasta nuclei in
collapsing stellar cores and neutron star crusts).  As the density increases, the shape of nuclei 
changes from spherical (SP) to cylindrical (C), slab-like (S), 
cylindrical-hole (CH), and spherical-hole (SH) shapes before the 
matter becomes uniform.  Whether the pasta structures exist or not depends 
strongly on $L$ \citep{OI2007}.  It is also suggested that 
observations of the crustal torsional oscillations enable us to 
extract the information about the pasta structures 
\citep{S2011,PP2016,SIO2017}. 
We also remark that elaborate dynamical model calculations
of the pasta structures have been done at conditions
marginally relevant for cold neutron stars 
\citep{Watanabe2003,SMF2011,DGL2012,SIMR2014,CSHB2015}.
The possibility that
more complicated structures than the above-mentioned shapes
may occur even at zero temperature has been suggested, but
in this work we assume that the density region where such
structures occur is negligible.

In spite of progress in theoretical researches, observational evidences 
for the presence of the pasta phases, let alone observational
constraints on the thickness of such phases, are basically lacking.
This is partly because the crust thickness is at most $\sim 10\%$ 
of the radius of a neutron star with canonical mass and partly because
even the star's mass and radius are hard to determine from observations.
On the other hand, the properties of the crust including the pasta 
structures could be important to the thermal evolution  
\citep{Newton2013,Hetal2015} and rotational evolution \citep{PVR2013}
of neutron stars.   Via observations associated with such 
evolutions, one could deduce the crustal properties.

In this paper, we focus on how the thickness of the neutron star crust 
and the pasta phases depends on the compactness and EOS parameters. 
For this purpose, we first obtain the equilibrium crust models
by numerically integrating the Tolman-Oppenheimer-Volkoff (TOV) equations 
together with an appropriate crust EOS, and then present a qualitative description 
of such thickness by simply combining the
TOV equations with the Gibbs-Duhem relation.  After that, we
construct fitting formulas for such thickness from the equilibrium
crust models obtained above.
We find that the thickness of the phase of 
spherical nuclei is strong function of the compactness, but 
relatively weak function of the EOS parameters, and 
confirm the known compactness dependence of the crust thickness.
Then, such thickness 
could be extracted from determination of the compactness within 
$\sim 10\%$ accuracy, independent of the EOS parameters.  We use units in 
which $c=G=1$, where $c$ and $G$ denote the speed of light and the 
gravitational constant, respectively.  Note that the compactness 
becomes a dimensionless parameter with the present units.

\section{Models for Neutron Star Crusts}
\label{sec:II}

We start with construction of equilibrium neutron star crusts.
For this purpose, it is convenient to write down the bulk energy 
per baryon of uniform nuclear matter at zero temperature 
in the vicinity of the saturation density, $n_0$, of symmetric 
nuclear matter as a function of baryon number density, $n_{\rm b}$, 
and neutron excess, $\alpha$, with four coefficients
($w_0$, $K_0$, $S_0$, and $L$) \citep{L1981}:
\begin{equation}
  w = w_0  + \frac{K_0}{18n_0^2}(n_{\rm b}-n_0)^2 + \left[S_0 
      + \frac{L}{3n_0}(n_{\rm b}-n_0)\right]\alpha^2.
  \label{eq:w}
\end{equation}
These coefficients and $n_0$ play the role of the 
saturation parameters, and each EOS has a corresponding set of the 
saturation parameters.  The saturation parameters have been gradually
well constrained from terrestrial nuclear experiments, while
among the five the incompressibility of symmetric nuclear matter, 
$K_0$, and the slope parameter, $L$, which are higher order 
coefficients with respect to density change from $n_0$, are relatively 
difficult to determine.  Thus, in describing the 
dependence of the crustal structure on the EOS of nuclear 
matter, we regard $K_0$ and $L$ as free parameters and fix 
the other saturation parameters ($n_0$, $w_0$, and $S_0$) in such a 
way as to reproduce empirical data for masses and charge radii of
stable nuclei.   In practice, we do so by using 
the phenomenological EOSs of nuclear matter constructed within the 
framework of the Thomas-Fermi theory by \cite{OI2003}.   The EOSs of 
beta-equilibrated, neutral matter in the crust were derived within the
same framework from the above EOS of nuclear matter by \cite{OI2007} 
(see also \cite{IO2014}).  Hereafter, we refer to such EOSs as 
OI-EOSs.  We remark that instead of $K_0$ and $L$,
the OI-EOSs are originally characterized by $K_0$ and $y$, where $y$ 
is defined as $y=-K_0S_0/(3n_0L)$, and that one can easily 
determine the value of $L$ for given $y$.  In Table \ref{tab:SH-density}, 
we show the sets of the saturation parameters that 
are adopted in this work.  Here, even extreme cases are effectively 
covered \citep{OI2003}, as compared to typical values obtained
from terrestrial experiments, e.g., $K_0=230\pm 40$ MeV \citep{KM2013} 
or $250< K_0< 315$ MeV \citep{SSM2014}, and $30 \lsim L \lsim 80$ 
MeV \citep{Newton2014}.

In order to construct neutron star models, one generally needs to 
prepare the EOS of matter in the star ranging from the star's 
surface down to center.  However, the EOS of matter in the core,
particularly in the density region higher than a few times 
normal nuclear density, is still uncertain.  To avoid such 
uncertainties, we construct the crust of a non-rotating neutron 
star with mass $M$ and radius $R$ by integrating the 
TOV equations from the star's surface 
inward down to the base of the crust, as in 
\cite{IS1997,SNIO2012,SIO2017}.   In this work, we focus on the 
stellar models with $1.4\le M/M_\odot \le 1.8$ and $10\le R \le 14$ km. 
Now, the crustal structure thus constructed is controlled by the  
four parameters, namely, the EOS parameters $K_0$ and $L$  and
the neutron star parameters $M$ and $R$.  We remark in passing
that the layer of the ocean is so thin that we can safely neglect the
the ocean thickness in calculating the crust thickness.

As mentioned above, the shapes of nuclei can change from spherical 
to cylindrical, slab-like, cylindrical-hole, and spherical-hole 
before the matter becomes uniform.  The densities at the respective
phase transitions depend on the saturation parameters \citep{OI2007}. 
In fact, for the EOS parameter sets adopted in this paper, 
the transition densities are listed in Table \ref{tab:SH-density}.

\begin{table}
\centering
\caption{
The SP-C, C-S, S-CH, CH-SH, and SH-U transition densities obtained for 
various sets of the EOS parameters, which are characterized by 
$K_0$ and $L$.  The asterisk at the value of $K_0$ denotes the EOS model by 
which some pasta phases are not predicted to appear.  That is, the values 
with *1, *2, and *3 denote the transition densities from 
cylindrical nuclei to uniform matter, from cylindrical-hole nuclei to uniform 
matter, and from spherical nuclei to uniform matter, respectively.
}
\begin{tabular}{ccccccccc}
\hline\hline
  $K_0$ (MeV) & $-y$ (MeV fm$^3$) & $L$ (MeV) & SP-C (fm$^{-3}$) & C-S (fm$^{-3}$) & S-CH (fm$^{-3}$) & CH-SH (fm$^{-3}$) & SH-U (fm$^{-3}$)  \\
\hline
  180 & 1800 & 5.7   & 0.06000 & 0.08665 & 0.12039 & 0.12925   & 0.13489     \\  
  180 & 600 & 17.5 & 0.05849 & 0.07986 & 0.09811 & 0.10206  &  0.10321     \\  
  180 & 350 & 31.0 & 0.05887 & 0.07629 & 0.08739 & 0.09000   & 0.09068     \\  
  180 & 220 & 52.2 & 0.06000 & 0.07186 & 0.07733 & 0.07885   & 0.07899     \\  
\hline
  230 & 1800 & 7.6   & 0.05816 & 0.08355 & 0.11440 & 0.12364   & 0.12736     \\  
  230 & 600 & 23.7 & 0.05957 & 0.07997 & 0.09515 & 0.09817  &  0.09866     \\  
  230 & 350 & 42.6 & 0.06238 & 0.07671 & 0.08411 & 0.08604   & 0.08637     \\  
  230 & 220 & 73.4 & 0.06421 & 0.07099 & 0.07284 & 0.07344   & 0.07345     \\  
\hline
  280 & 1800 & 9.6  & 0.05747 & 0.08224 & 0.11106 & 0.11793  &  0.12286   \\ 
  280 & 600 & 30.1 & 0.06218 & 0.08108 & 0.09371 & 0.09577 &  0.09623    \\ 
  280 & 350 & 54.9 & 0.06638 & 0.07743 & 0.08187 & 0.08314 &   0.08331   \\ 
  $^*$280 & 220 & 97.5 & 0.06678 & --- & ---  & ---  &  0.06887$^{*1}$  \\ 
\hline
  360 & 1800 & 12.8 & 0.05777 & 0.08217 & 0.10892 & 0.11477   & 0.11812     \\  
  360 & 600 & 40.9 & 0.06743 & 0.08318 & 0.09197 & 0.09379  &  0.09414     \\  
  $^*$360 & 350 & 76.4   & 0.07239 & 0.07797 & 0.07890 & --- & 0.07918$^{*2}$     \\  
  $^*$360 & 220 & 146.1 & --- & --- & --- & ---   & 0.06680$^{*3}$     \\  
\hline\hline
\end{tabular}
\label{tab:SH-density}
\end{table}

\begin{table}
\centering
\caption{
Same as Table 1 but for the neutron chemical potential, including the 
rest mass, at the phase transitions.
}
\begin{tabular}{ccccccccc}
\hline\hline
  $K_0$ (MeV) & $-y$ (MeV fm$^3$) & $L$ (MeV) & SP-C (MeV) & C-S (MeV) & S-CH (MeV) & CH-SH (MeV) & SH-U (MeV)  \\
\hline
  180 & 1800 & 5.7 & 951.041 & 952.272 & 952.918  &  953.151  &  953.291    \\  
  180 & 600 & 17.5 & 950.667 & 952.004 & 952.971  &  953.175  &  953.232    \\  
  180 & 350 & 31.0 & 950.318 & 951.712 & 952.434  &  952.680  &  952.747    \\  
  180 & 220 & 52.2 & 949.839 & 951.114 & 951.611  &  951.812  &   951.831   \\  
\hline
  230 & 1800 & 7.6 & 950.915 & 952.195 & 953.119 &  953.208  &   953.319   \\  
  230 & 600 & 23.7 & 950.583 & 952.069 & 952.915 &  953.176  &   953.219   \\  
  230 & 350 & 42.6 & 950.307 & 951.779 & 952.441 &  952.689  &   952.732   \\  
  230 & 220 & 73.4 & 949.931 & 950.949 & 951.196 &  951.304  &   951.305   \\  
\hline
  280 & 1800 & 9.6 & 950.837 & 952.169 & 952.904 &  953.168 &    953.400  \\ 
  280 & 600 & 30.1 & 950.625 & 952.224 & 953.286 &  953.459 &    953.496  \\ 
  280 & 350 & 54.9 & 950.552 & 952.050 & 952.571 &  952.780 &    952.807  \\ 
  $^*$280 & 220 & 97.5 & 950.344 & --- & ---  & ---  &  950.743$^{*1}$  \\ 
\hline
  360 & 1800 & 12.8 & 950.812 & 952.244 & 953.307 &  953.397  &   953.611   \\  
  360 & 600 & 40.9   & 950.968 & 952.756 & 953.806 &  954.014  &   954.051   \\  
  $^*$360 & 350 & 76.4   &  951.535  &  951.929  & 951.995 & --- & 952.014$^{*2}$     \\  
  $^*$360 & 220 & 146.1 & --- & --- & --- & ---   & 951.277$^{*3}$     \\  
\hline\hline
\end{tabular}
\label{tab:SH-chempot}
\end{table}

\section{Formulas for the thickness of the pasta phases 
and the whole crust}
\label{sec:III}

In this section, we derive fitting formulas for the thickness of the whole 
and parts of the neutron star crust constructed in the previous section.  
Before going into details, we give a qualitative description of such thickness
by combining the TOV equations with the Gibbs-Duhem relation as 
\begin{equation}
\Delta R_{AB}\simeq \frac{(R-C\Delta R_{AB})^2 [1-2M/(R-C\Delta R_{AB})]}{m_n M}(\mu_B-\mu_A),
\label{eq:mvt}
\end{equation}
where $\Delta R_{AB}$ is the crust thickness between two radii $B$ (lower) and 
$A$ (upper), $m_n$ is the neutron rest mass, $\mu_A$ and $\mu_B$ are the 
neutron chemical potentials including $m_n$ at the radii $A$ and $B$, and 
$C$ is a constant that comes from the mean-value theorem and satisfies 
$0\leq C\leq1$.  In Eq.\ (\ref{eq:mvt}),  the pressure is ignored as compared
with the mass density, which is in turn approximated as $m_n$ times baryon 
density.  We also assume that the mass of the crust is negligibly small
compared with $M$.  Instead of solving Eq.\ (\ref{eq:mvt}) with respect to 
$\Delta R_{AB}$, we can obtain an approximate solution by setting $C=0$ as 
\citep{PR1995}
\begin{equation}
\Delta R_{AB}\simeq \frac{R^2 (1-2M/R)}{m_n M}(\mu_B-\mu_A).
\label{eq:PR1995}
\end{equation}
We find from this expression that the ratio of the thickness $\Delta R_{AB}$
to $R$ is controlled by the compactness $M/R$ via the factor $R/M(1-2M/R)$ and 
also by the neutron chemical potential difference $\mu_B-\mu_A$.  Eventually,
by comparing with numerical results that will be given below, expression 
(\ref{eq:PR1995}) turns out to be successful in reproducing the $L$, $K_0$, 
and $M/R$ dependence of the thickness of the crust and pasta phases 
qualitatively, while deviations from the numerical results are at most
a factor of two.  Incidentally, all the values of $\mu_A$ and $\mu_B$ 
to be given in Eq.\ (\ref{eq:PR1995}) are listed in Table \ref{tab:SH-chempot}, 
except in the case of the thickness of the crust and the spherical phase
in which $\mu_A$ is set to $m_n$.
We remark that \cite{ZH2011,ZFH2017} have also derived the 
approximate relation between the crust thickness and the stellar 
compactness by ignoring the pressure correction terms in the TOV 
equations as we do in Eq. (\ref{eq:mvt}).

To examine how the chemical potential difference $\mu_B-\mu_A$ depends 
on the EOS parameters, it is convenient to obtain the expansion form of the 
neutron chemical potential from Eq.\ (\ref{eq:w}) as
\begin{equation}
\mu_n=m_n+w_0+S_0\alpha(2-\alpha)+
        \frac{n_b-n_0}{18n_0}\left(12L\alpha+K_0\frac{3n_b-n_0}{n_0}\right)
        +\frac{L}{3}\alpha^2,
\label{eq:munexp}
\end{equation}
This expression clearly shows the $L$ and $K_0$ contributions to $\mu_n$.
Although Eq.\ (\ref{eq:munexp}) is strictly valid near $n_b=n_0$ and $\alpha=0$,
it is instructive to extrapolate it to the regime of $n_b$ and $\alpha$ relevant
to the deepest region of the crust, i.e., subnuclear densities and extremely 
large neutron excess.  Note that all the transition densities listed in Table 
\ref{tab:SH-density} (see also Fig.\ 1 of \cite{SIO2017}) lies between $n_0/3$ 
and $n_0$ and that a gas of dripped neutrons occupy more than $\sim$70\%
of the nucleons.  Thus, we can simply set $\alpha=1$, which gives rise to
$w_0+S_0+L/3$ as a constant part of $\mu_n-m_n$.  This part is typically
of order 10--40 MeV.  The remaining density-dependent part, which is 
negative and roughly of order 10 MeV, controls the EOS dependence of the 
thickness of each pasta-like nonspherical phase because the constant part 
is essentially cancelled in the difference $\mu_B-\mu_A$.  This EOS dependence 
is complicated by the fact that the transition densities themselves depend on 
$L$ and $K_0$ as in Table \ref{tab:SH-density}.  Anyway, according to Eq.\ 
(\ref{eq:munexp}), the $L$ dependence of $\mu_B-\mu_A$ is dominant over the 
$K_0$ dependence, which will play a role in parametrizing the thickness of
the pasta phases as a function of $M/R$, $L$, and $K_0$.

The thickness of the spherical phase needs to be examined separately.  In this
case, the constant part $w_0+S_0+L/3$ contributes to the $L$ dependence of 
$\mu_B-\mu_A$ (here, $\mu_A=m_n$) in such a way that the $L$ dependence that 
comes from the density-dependent part as shown above is weakened.  Since
the SP-C transition density is about $n_0/3$ and almost independent of $L$ and 
$K_0$, furthermore, the thickness of the spherical phase is expected to depend 
only weakly on $L$ and $K_0$.  This expectation looks consistent with the 
behavior of $\mu_n$ that can be seen from Table \ref{tab:SH-chempot}.  We 
remark that the thickness of the whole crust, which is dominated by the 
thickness of the spherical phase, has a similarly weak EOS dependence.

As a typical thickness of the whole crust, we will thus use
\begin{equation}
  \frac{\Delta R}{R} \simeq 2.1\times 10^{-2} \frac{R}{M} 
                     \left(1-\frac{2M}{R}\right),  
  \label{eq:typ}
\end{equation}
which is based on Eq.\ (\ref{eq:mvt}) and is independent of $L$ and 
$K_0$.  Here, the factor $2.1\times 10^{-2}$ is slightly different from the 
factor $2.57\times 10^{-2}$ that were obtained by calculating the factor 
$\mu_B/m_n-1$ in Eq.\ (\ref{eq:PR1995}) from the EOS of FPS \citep{RP1994}.  
Note that the factor $2.1\times 10^{-2}$ effectively allows for nonzero $C$ 
in contrast to the factor $2.57\times 10^{-2}$.  
We remark that \cite{ZH2011,ZFH2017} have also
indicated the strong compactness dependence of the relative crust thickness, 
while studying the effects of accretion and rotation.
In the present work, we try 
to derive a fitting formula for the thickness of the whole crust by 
including the detailed $L$ and $K_0$ dependence.  Before doing so, in the 
following subsections, we consider the thickness of the SP phase and of each
pasta phase.

\subsection{Phase of spherical (SP) nuclei}
\label{sec:III-A}
\begin{figure*}
\begin{center}
\begin{tabular}{cc}
\includegraphics[scale=0.5]{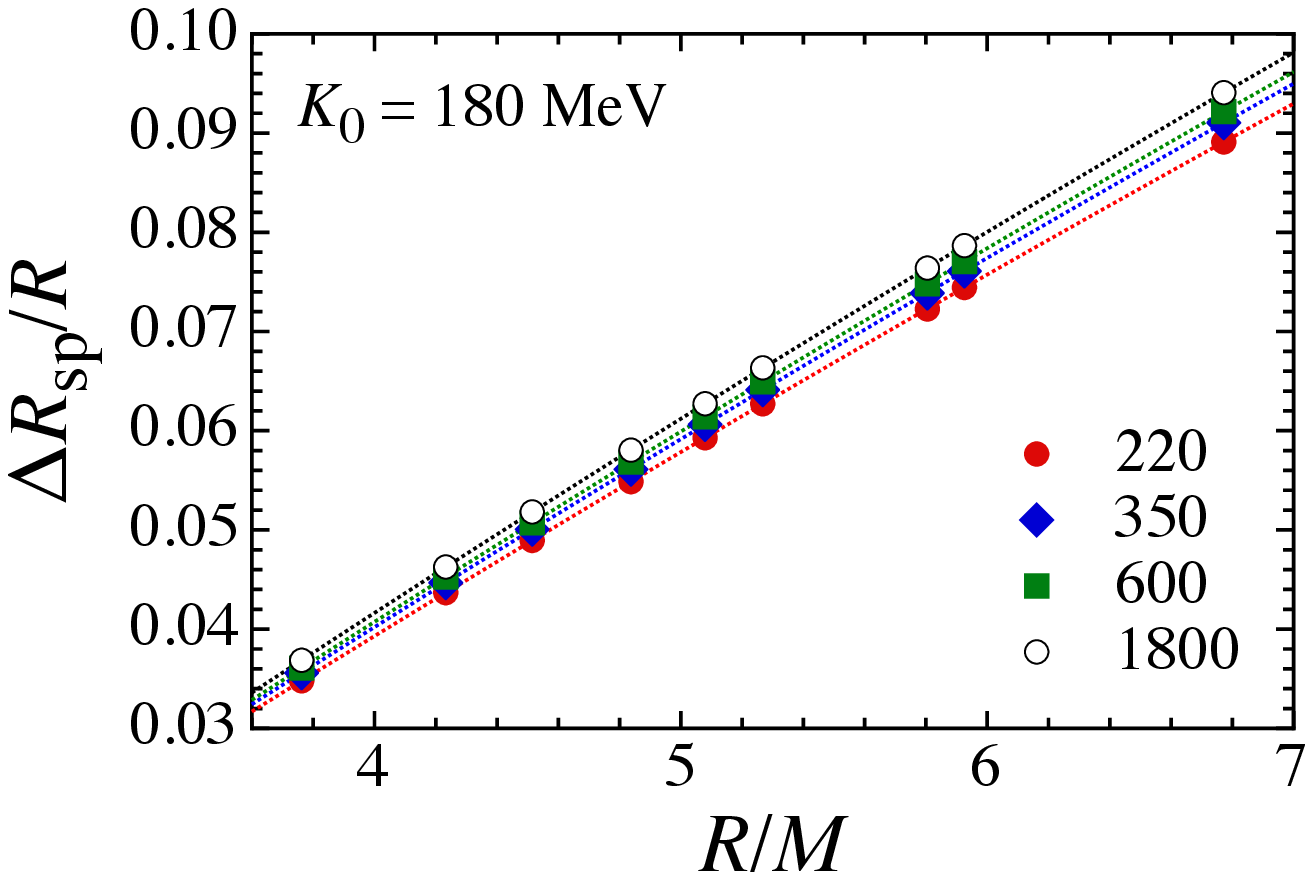} &
\includegraphics[scale=0.5]{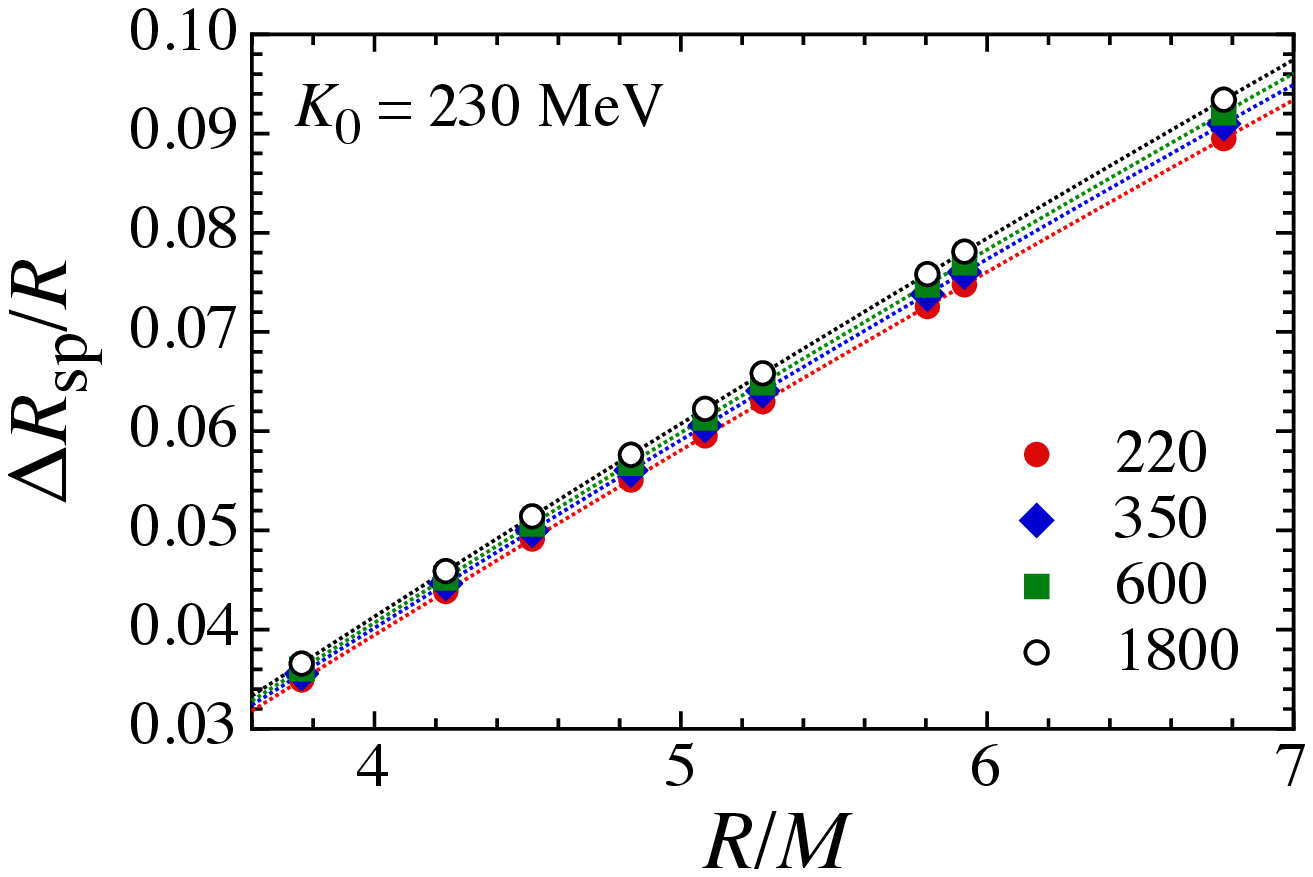} \\
\includegraphics[scale=0.5]{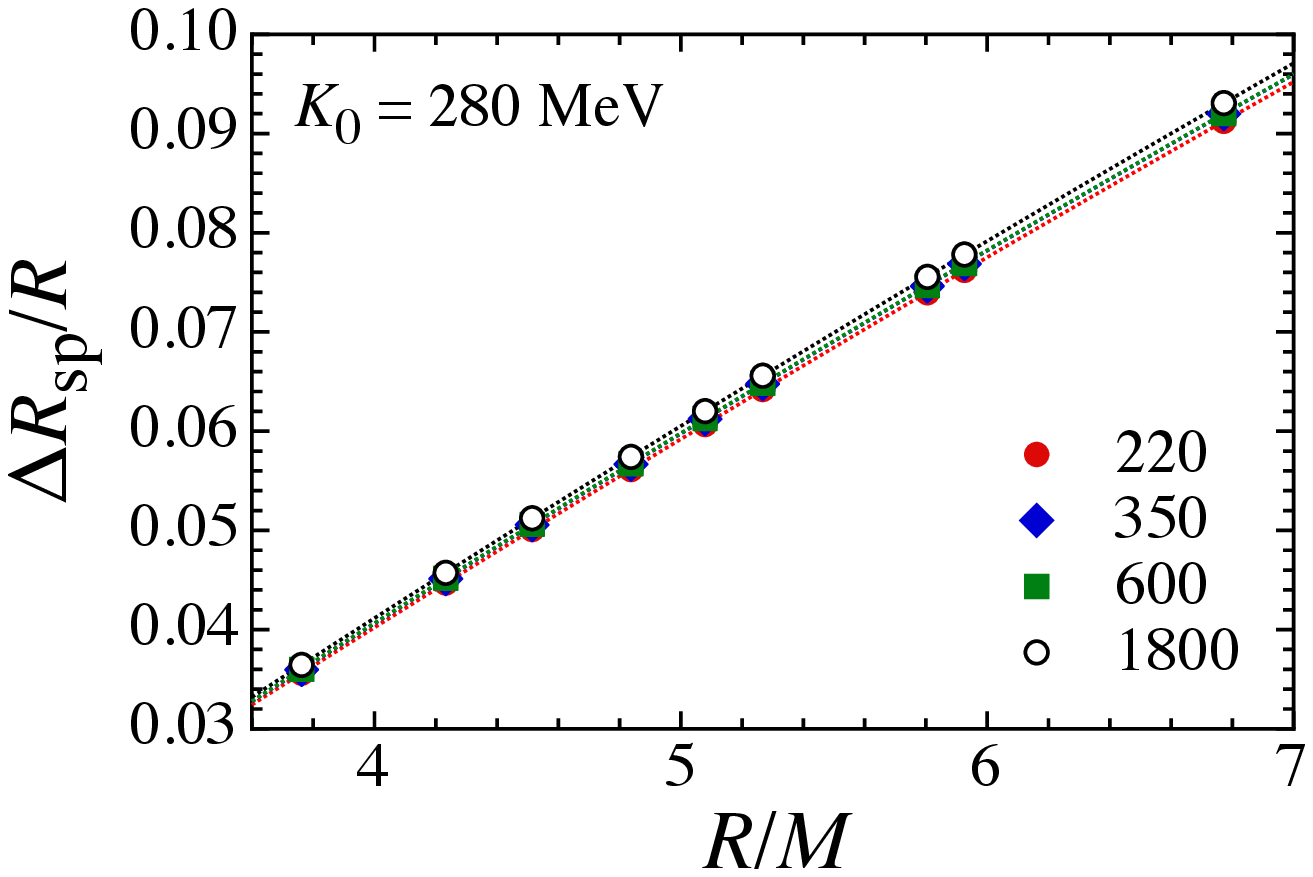} &
\includegraphics[scale=0.5]{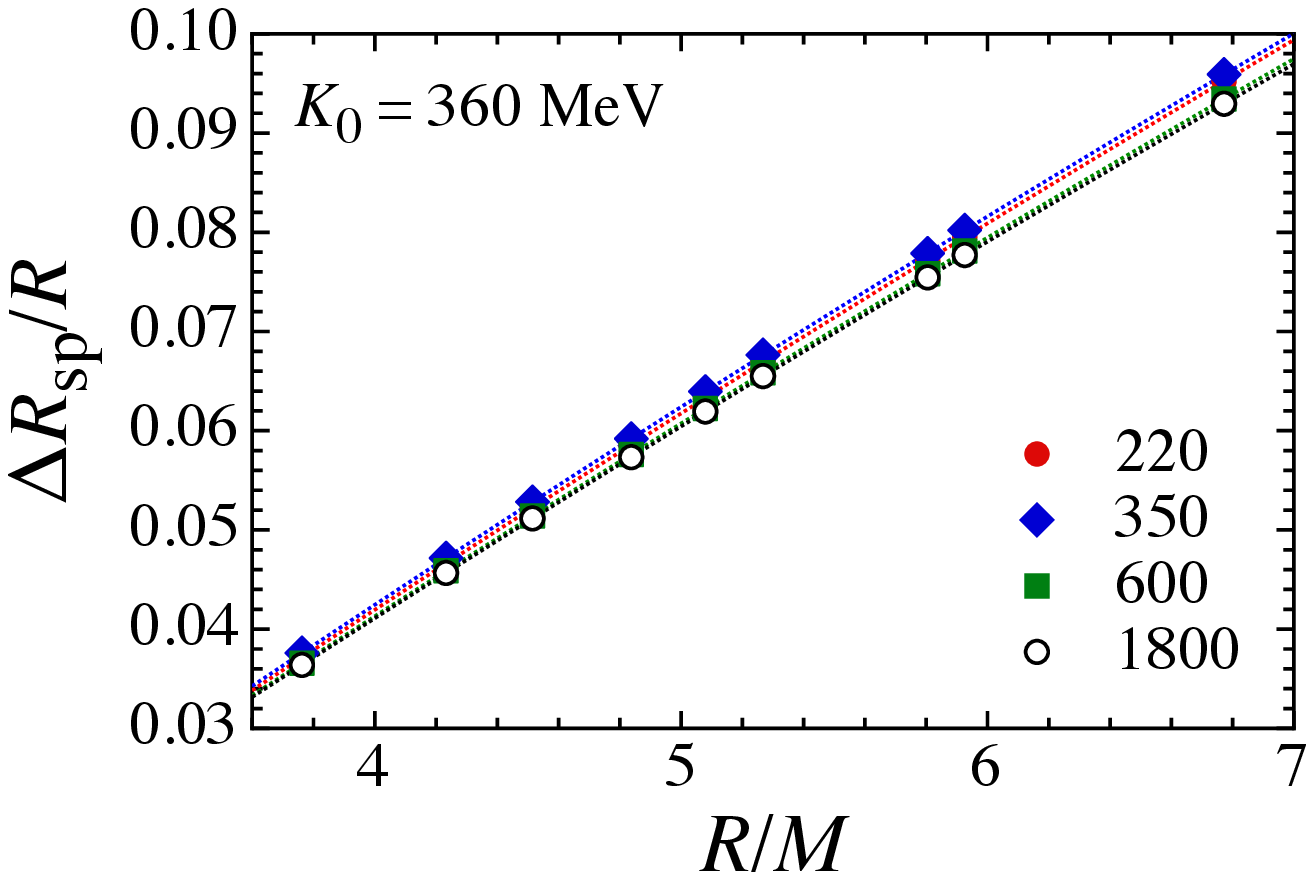} 
\end{tabular}
\end{center}
\caption{
The ratio of the thickness of the phase composed of spherical nuclei 
to the star's radius, which was obtained from various 
OI-EOSs, as a function of the inverse of the
compactness $M/R$.  The upper-left, upper-right,
lower-left, and lower-right panels correspond to the cases 
of $K_0=180$, 230, 280, and 360 MeV, while the solid-circles,
solid-diamonds, solid-squares, and open-circles in each panel 
are the results obtained for $-y=220$, $350$, $600$, and 
$1800$ MeV fm$^3$.  The dotted lines denote the fitting formula 
given by Eq.\ (\ref{eq:dRsp-MR}).
}
\label{fig:dRsp-MR}
\end{figure*}

By using the neutron star crusts constructed in Sec.\ \ref{sec:II}
for nine stellar models with the combinations of three different 
masses ($M/M_\odot=1.4$, 1.6, and 1.8) and three different radii ($R=10$, 
12, and 14 km), let us now examine the compactness and EOS
dependence of the thickness, $\Delta R_{\rm sp}$, of the phase composed of 
the spherical nuclei.  In Fig.\ \ref{fig:dRsp-MR}, we show the ratio of 
$\Delta R_{\rm sp}$ to $R$ as a function of $R/M$, which is the reciprocal 
of the stellar compactness, for various sets of $L$ and $K_0$. 
From this figure, we find that $\Delta R_{\rm sp}/R$ can be 
well expressed as a function of $R/M$ for each set of the EOS 
parameters:
\begin{equation}
  \frac{\Delta R_{\rm sp}}{R} = -\alpha_1^{\rm sp} \left(\frac{R}{M}\right)^2 + \alpha_2^{\rm sp} \left(\frac{R}{M}\right) - \alpha_3^{\rm sp}, 
    \label{eq:dRsp-MR}
\end{equation}
where $\alpha_1^{\rm sp}$, $\alpha_2^{\rm sp}$, and $\alpha_3^{\rm sp}$ are 
positive dimensionless adjustable coefficients that depend
on ($L$, $K_0$).  Note that this form arises from Eq.\ (\ref{eq:mvt}) 
in which the parameter $C$ is taken to be order unity.  We can then expect
$\alpha_1^{\rm sp}$ to be small compared with $\alpha_2^{\rm sp}$ and 
$\alpha_3^{\rm sp}$.  In Fig.\ \ref{fig:dRsp-MR}, we can confirm that
expression (\ref{eq:dRsp-MR}) does accurately reproduce $\Delta R_{\rm sp}/R$ 
for each set of the EOS parameters.  In addition, one can observe 
that $\Delta R_{\rm sp}/R$ strongly depends on the stellar compactness, 
while the dependence on the EOS parameters is relatively weak, 
as expected from the above-mentioned arguments.  That is, 
one can deduce the value of $\Delta R_{\rm sp}/R$ 
once the stellar compactness is observationally determined.

\begin{figure*}
\begin{center}
\begin{tabular}{ccc}
\includegraphics[scale=0.4]{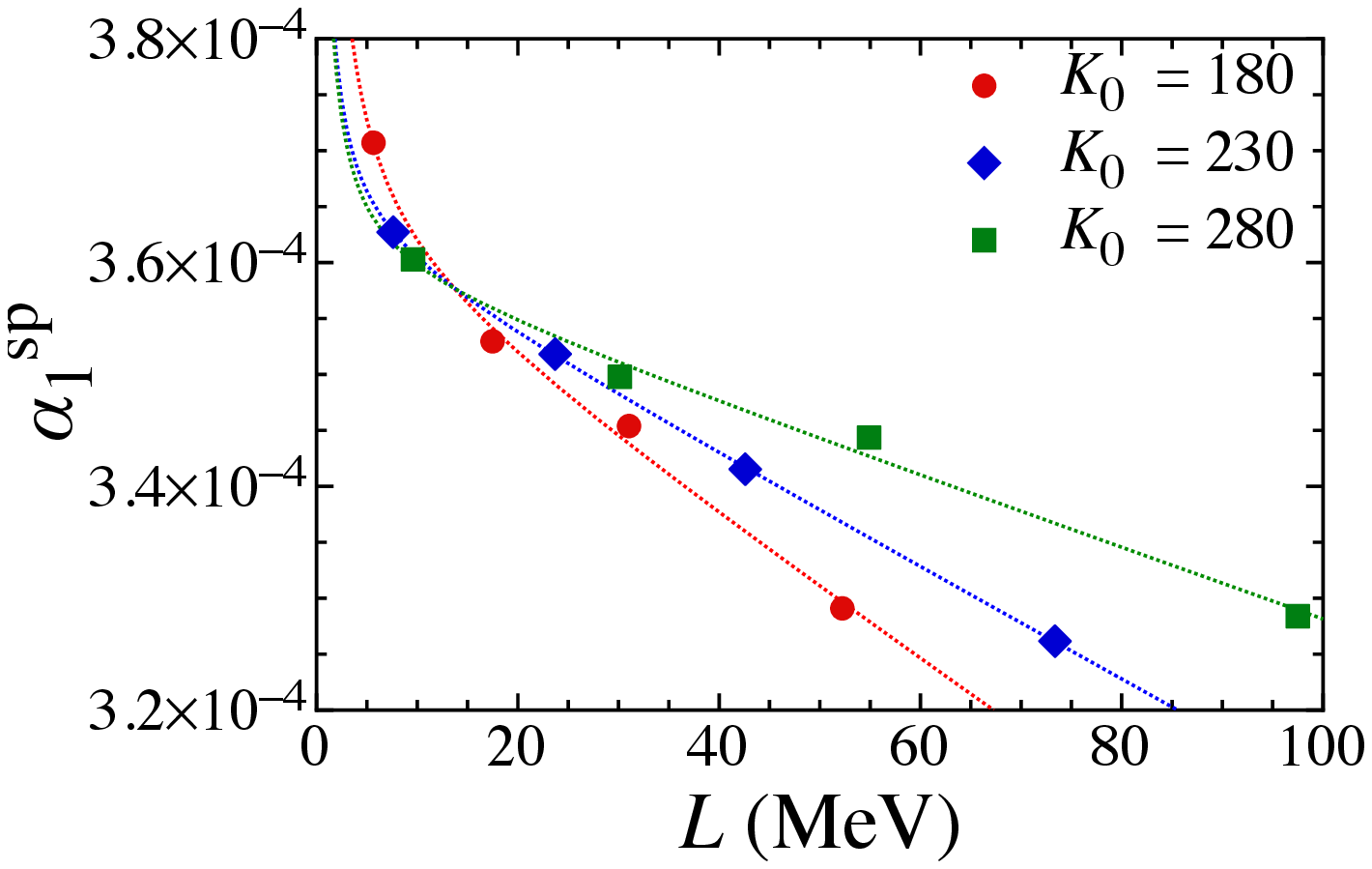} &
\includegraphics[scale=0.4]{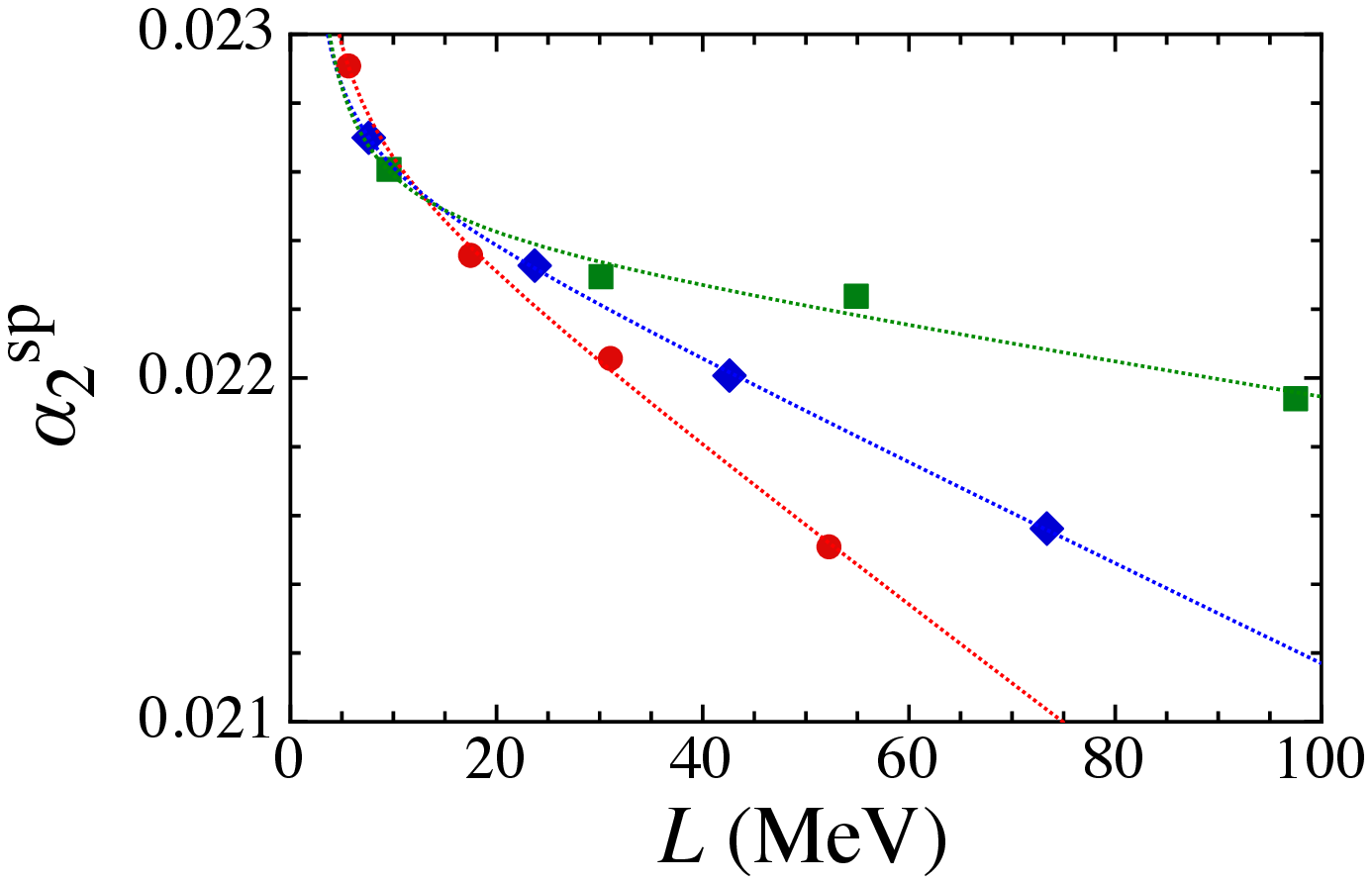} &
\includegraphics[scale=0.4]{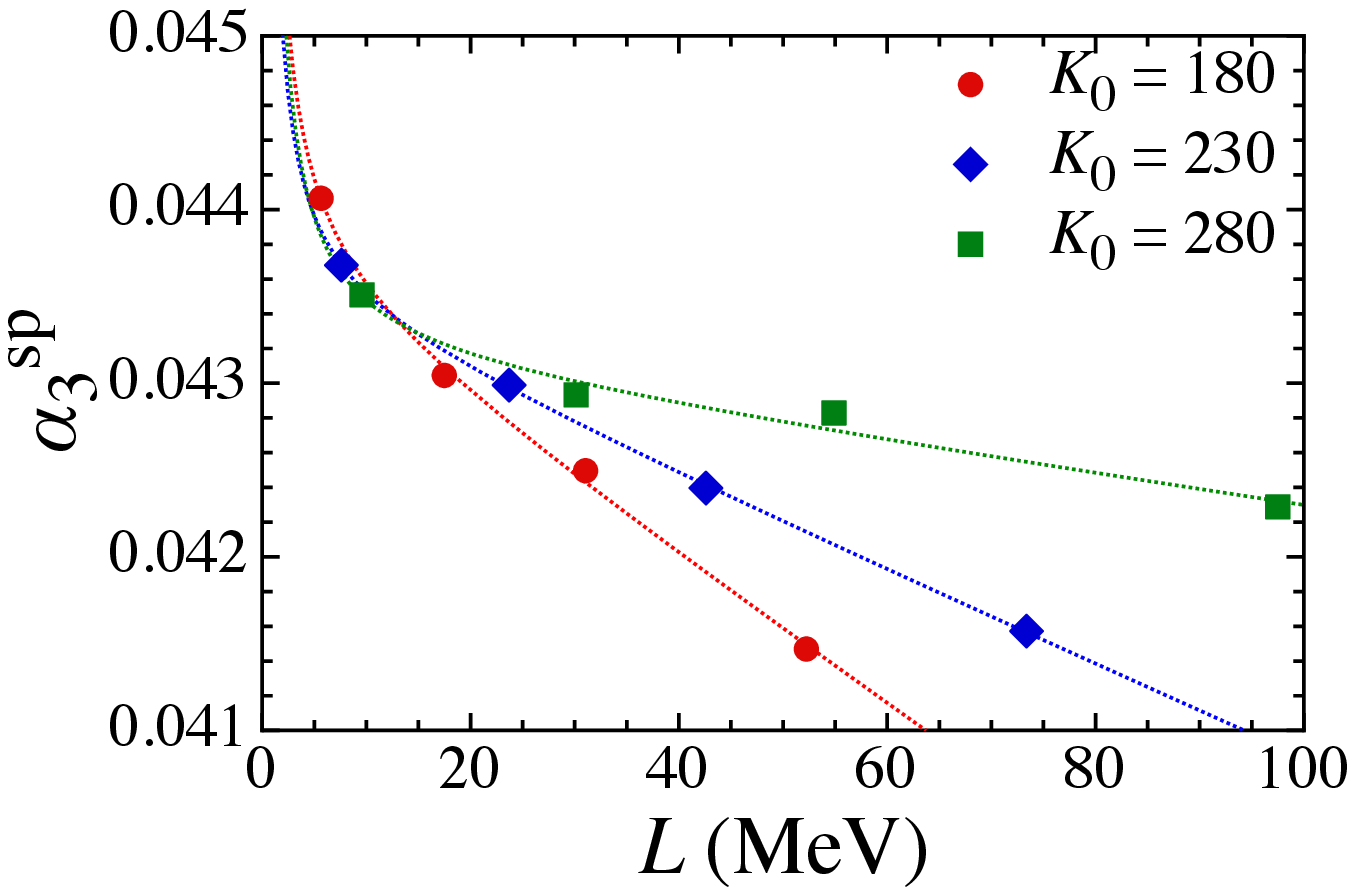} 
\end{tabular}
\end{center}
\caption{
The coefficients in Eq.\ (\ref{eq:dRsp-MR}), shown 
as a function of $L$. The circles, diamonds, and squares 
are the results of $K_0=180$, 230, and 280 MeV. 
The dotted lines in each panel denote the fitting formula 
given by Eqs.\ (\ref{eq:dRsp-a1})--(\ref{eq:dRsp-a3}).
}
\label{fig:asp-L}
\end{figure*}

We then move on to express the coefficients in Eq.\ 
(\ref{eq:dRsp-MR}) as a function of the EOS parameters 
$(L, K_0)$.  In Fig.\ \ref{fig:asp-L} we plot the 
values of $\alpha_i^{\rm sp}$ with $i=1$, 2, and 3, 
which were obtained by fitting for several sets of $K_0$ 
and $L$. From this figure, we find that $\alpha_i^{\rm sp}$ 
with $i=1$, 2, and 3 can be expressed as a function of 
$L$ for $K_0=180$, 230, and 280 MeV by
\begin{eqnarray}
  \alpha_1^{\rm sp} &=& \beta_{11}^{\rm sp} \left(\frac{L}{60\ {\rm MeV}}\right)^{-1} 
       + \beta_{12}^{\rm sp} - \beta_{13}^{\rm sp}\left(\frac{L}{60\ {\rm MeV}}\right),
       \label{eq:dRsp-a1} \\
  \alpha_2^{\rm sp} &=& \beta_{21}^{\rm sp} \left(\frac{L}{60\ {\rm MeV}}\right)^{-1} 
       + \beta_{22}^{\rm sp} - \beta_{23}^{\rm sp}\left(\frac{L}{60\ {\rm MeV}}\right),
       \label{eq:dRsp-a2} \\  
  \alpha_3^{\rm sp} &=& \beta_{31}^{\rm sp} \left(\frac{L}{60\ {\rm MeV}}\right)^{-1} 
       + \beta_{32}^{\rm sp} - \beta_{33}^{\rm sp}\left(\frac{L}{60\ {\rm MeV}}\right), 
       \label{eq:dRsp-a3}  
\end{eqnarray}
where $\beta_{ij}^{\rm sp}$ with $i=1$, 2, 3 and $j=1$, 2, 3 are 
positive dimensionless fitting parameters that depend
on $K_0$.  Figure \ref{fig:asp-L} shows that 
expressions (\ref{eq:dRsp-a1})--(\ref{eq:dRsp-a3}) accurately 
reproduce the $L$ dependence of the coefficients in
Eq.\ (\ref{eq:dRsp-MR}) for $K_0=180$, 230, 280 MeV. 
We remark that this is not the case with $K_0=360$ MeV, but this 
value of $K_0$ is obviously beyond the constraint from the 
terrestrial experiments (e.g., \cite{KM2013,SSM2014}).

\begin{figure}
\begin{center}
\includegraphics[scale=0.5]{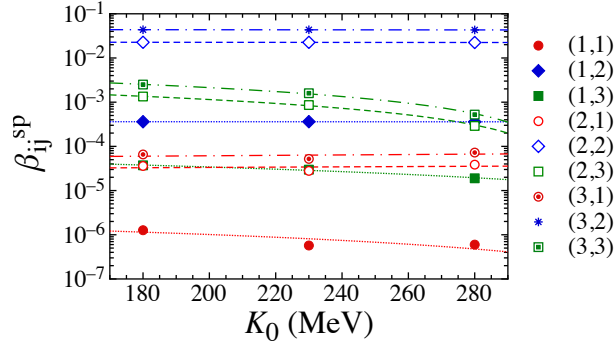}
\end{center}
\caption{
The coefficients in Eqs.\ (\ref{eq:dRsp-a1})--(\ref{eq:dRsp-a3})
plotted as a function of $K_0$, where the labels of $(i,j)$ for 
$i=1,2,3$ and $j=1,2,3$ denote the subscript in 
$\beta_{ij}^{\rm sp}$.  The dotted, dashed, and dot-dashed lines 
denote the fitting formula for $i=1$, 2, and 3, respectively.
}
\label{fig:bsp-K0}
\end{figure}

Finally, we construct the fitting formula for the 
coefficients in Eqs.\ (\ref{eq:dRsp-a1})--(\ref{eq:dRsp-a3}) 
as a function of $K_0$.  In Fig.\ \ref{fig:bsp-K0}, we 
plot the values of $\beta_{ij}^{\rm sp}$ obtained
by fitting for $K_0=180, 230, 280$ MeV as shown in Fig.\
\ref{fig:asp-L}.  From Fig.\ \ref{fig:bsp-K0}, we find that the values 
of $\beta_{ij}^{\rm sp}$ can then be fitted as a linear 
function of $K_0$:
\begin{eqnarray}
  \beta_{11}^{\rm sp} &=& \left[ 2.3515   -  1.5395 \left(\frac{K_0}{230\ {\rm MeV}}\right)\right] \times 10^{-6}, 
      \label{eq:dRsp-b11} \\
  \beta_{12}^{\rm sp} &=&  \left[ 3.6353  - 0.02852 \left(\frac{K_0}{230\ {\rm MeV}}\right)\right] \times 10^{-4},
     \label{eq:dRsp-b12} \\
  \beta_{13}^{\rm sp} &=&  \left[ 7.0703  -  4.2055\left(\frac{K_0}{230\ {\rm MeV}}\right)\right] \times 10^{-5},
    \label{eq:dRsp-b13} \\
  \beta_{21}^{\rm sp} &=&  \left[ 2.8280   +   0.5881\left(\frac{K_0}{230\ {\rm MeV}}\right)\right] \times 10^{-5},
    \label{eq:dRsp-b21} \\
  \beta_{22}^{\rm sp} &=& \left[ 2.3107  -  0.05589 \left(\frac{K_0}{230\ {\rm MeV}}\right)\right] \times 10^{-2},
     \label{eq:dRsp-b22} \\
  \beta_{23}^{\rm sp} &=&  \left[ 3.2551   -  2.4233 \left(\frac{K_0}{230\ {\rm MeV}}\right)\right] \times 10^{-3},
    \label{eq:dRsp-b23} \\
  \beta_{31}^{\rm sp} &=& \left[ 0.4770   +   0.1564\left(\frac{K_0}{230\ {\rm MeV}}\right)\right] \times 10^{-4},
    \label{eq:dRsp-b31} \\
  \beta_{32}^{\rm sp} &=& \left[ 4.4481 -   0.1081\left(\frac{K_0}{230\ {\rm MeV}}\right)\right] \times 10^{-2},
    \label{eq:dRsp-b32} \\
  \beta_{33}^{\rm sp} &=& \left[ 6.0989  -   4.5574\left(\frac{K_0}{230\ {\rm MeV}}\right)\right] \times 10^{-3}.
    \label{eq:dRsp-b33} 
\end{eqnarray}
Now, we obtain a complete set of the fitting formulas 
(\ref{eq:dRsp-MR})--(\ref{eq:dRsp-b33}), which well reproduces
the calculated values of $\Delta R_{\rm sp}/R$ for various
combinations of $R/M$, $L$, and $K_0$.  Note that 
applicability of these formulas is limited to
the range of $180\lsim K_0\lsim 280$ MeV.

\subsection{Phase of cylindrical (C) nuclei}
\label{sec:III-B}

\begin{figure*}
\begin{center}
\begin{tabular}{cc}
\includegraphics[scale=0.5]{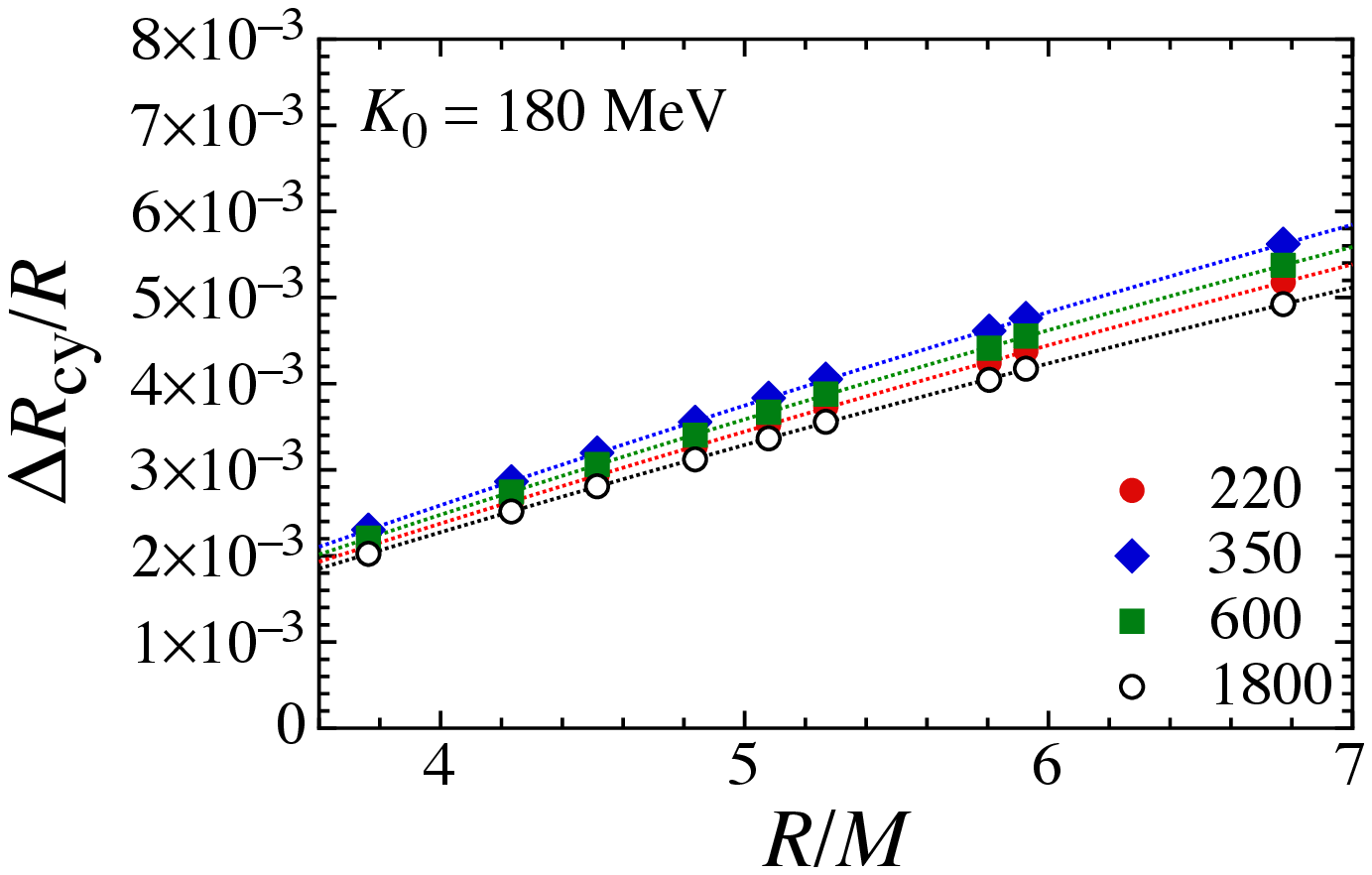} &
\includegraphics[scale=0.5]{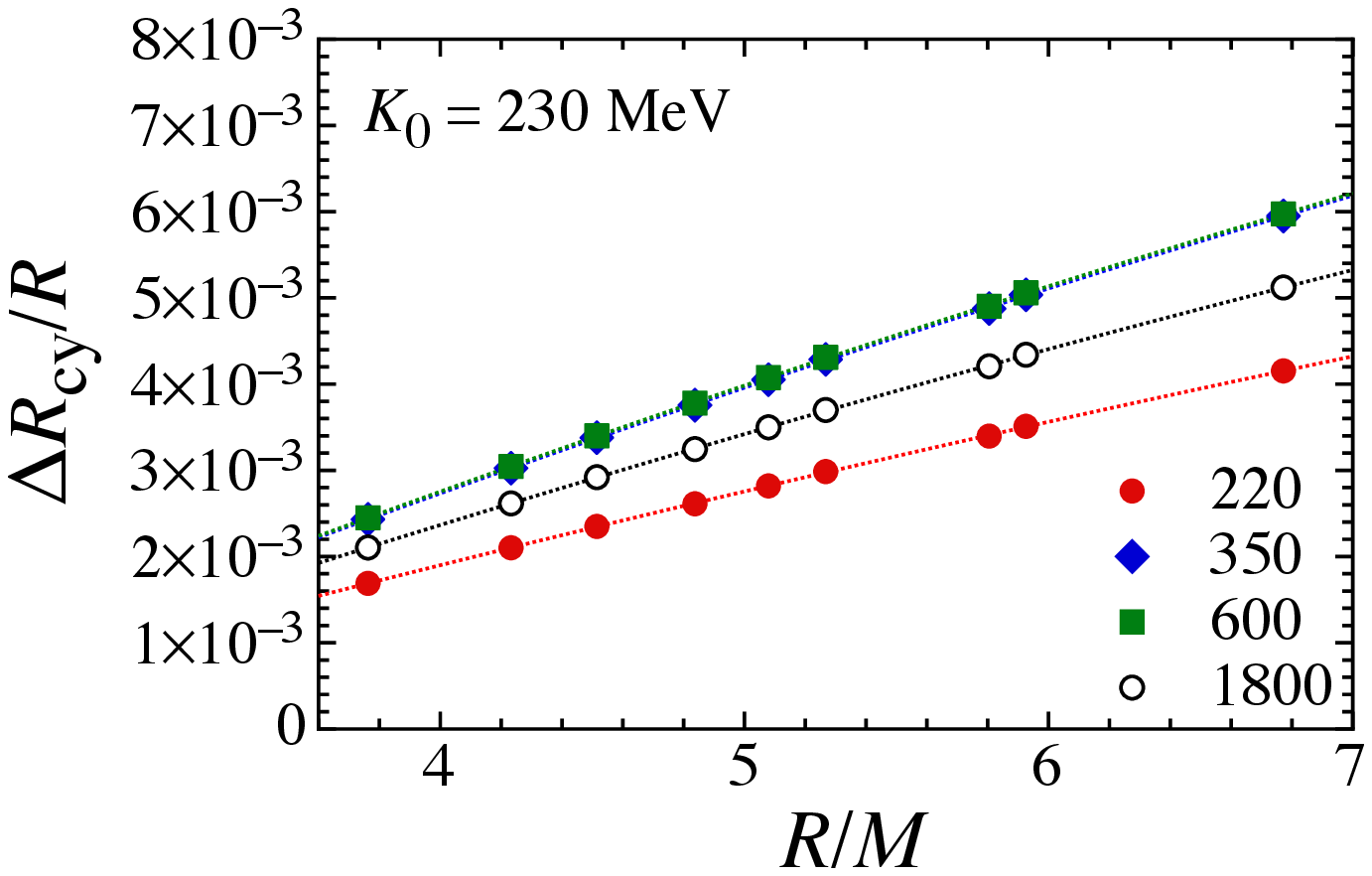} \\
\includegraphics[scale=0.5]{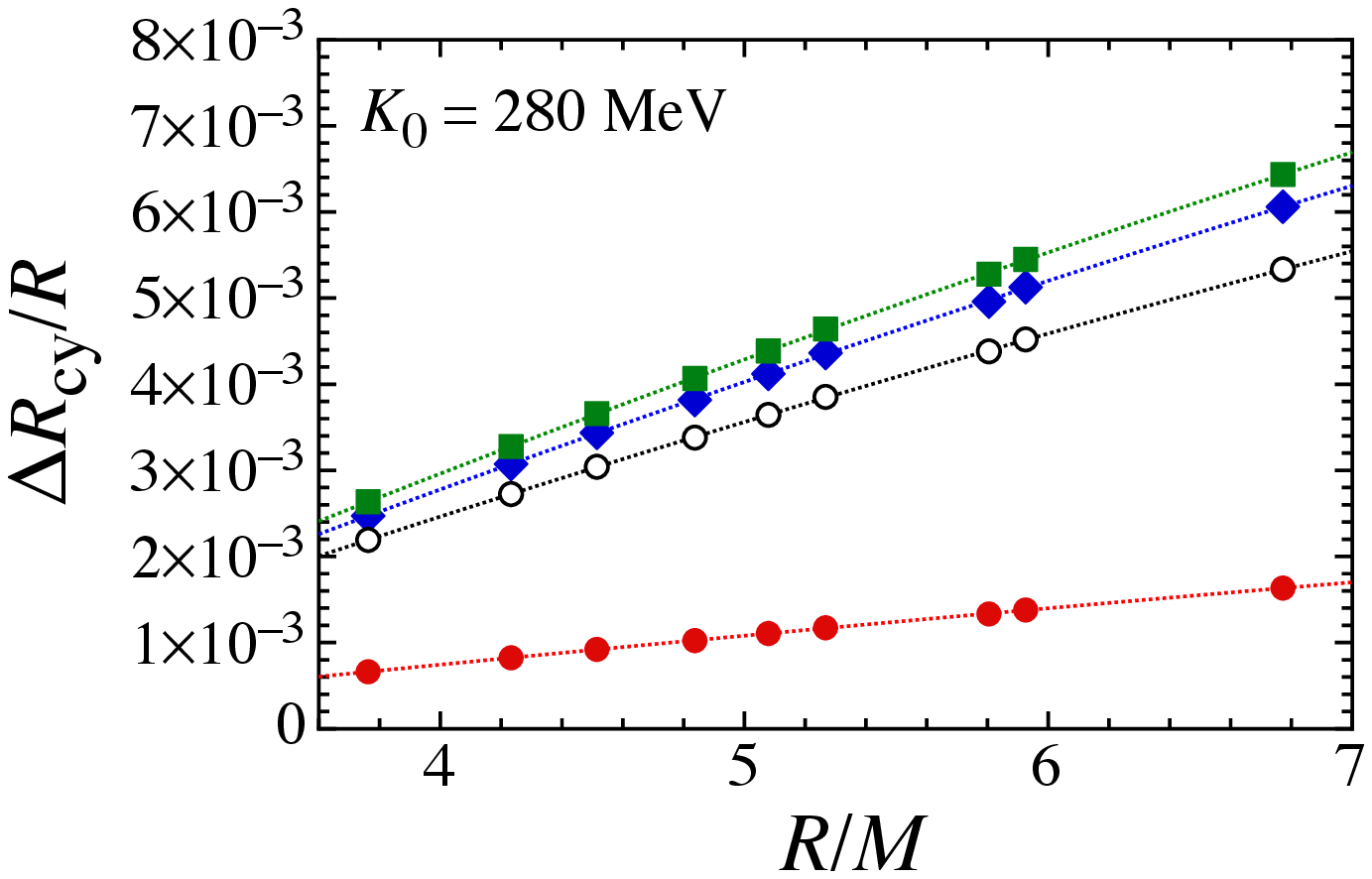} &
\includegraphics[scale=0.5]{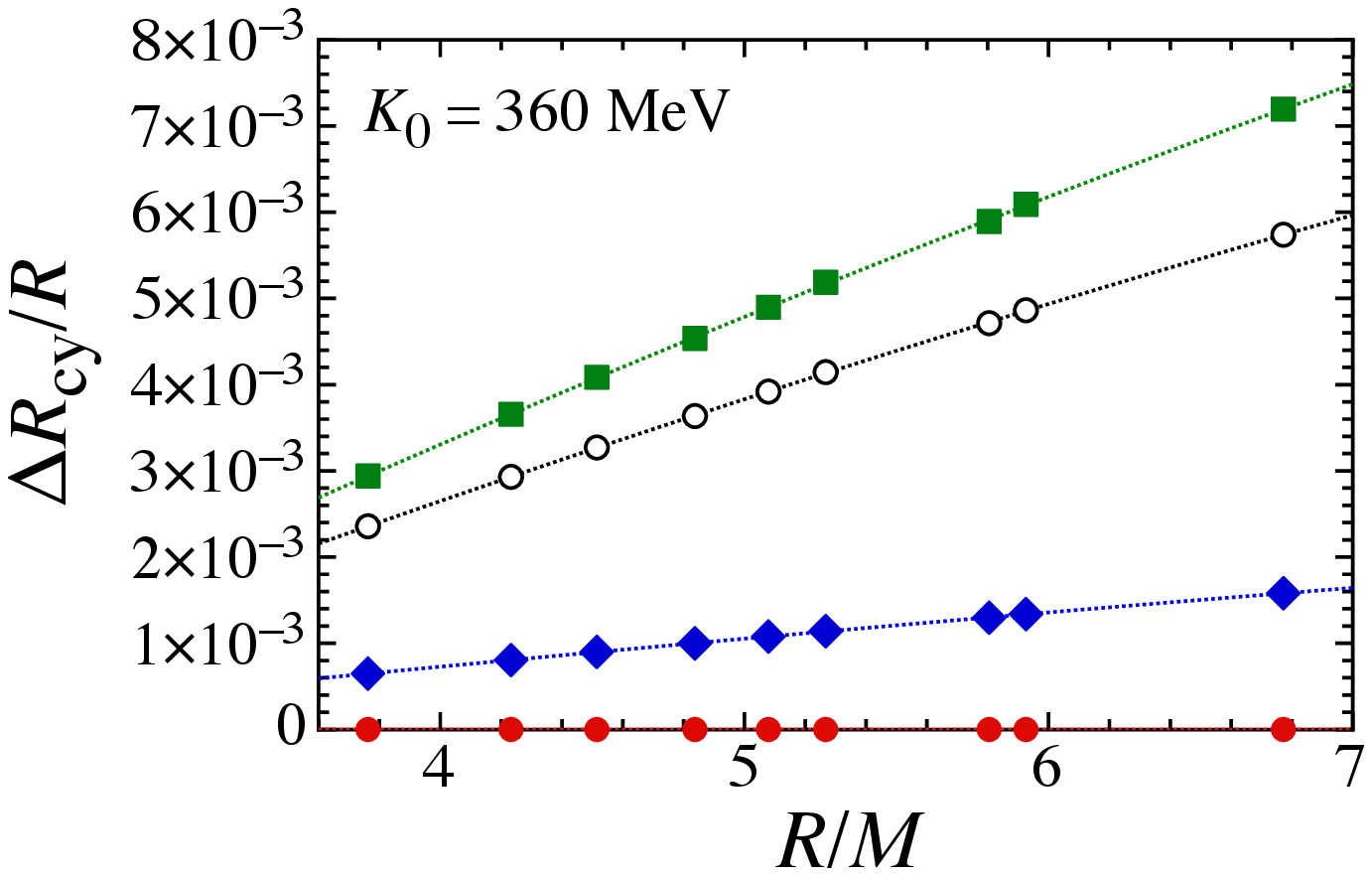} 
\end{tabular}
\end{center}
\caption{
Same as Fig.\ \ref{fig:dRsp-MR}, but for the ratio of the thickness of the phase
composed of cylindrical nuclei to the star's radius.  The dotted lines 
denote the fitting formula given by Eq.\ (\ref{eq:dRcy-MR}).
}
\label{fig:dRcy-MR}
\end{figure*}

Next, we turn to the thickness of the phase composed of cylindrical 
nuclei, $\Delta R_{\rm cy}$.   We find again that $\Delta R_{\rm cy}/R$ can be 
well expressed as a function of $R/M$ for each set of the EOS parameters,
as shown in Fig.\ \ref{fig:dRcy-MR}, i.e., we can derive the fitting formula 
for $\Delta R_{\rm cy}/R$ by
\begin{equation}
  \frac{\Delta R_{\rm cy}}{R} = -\alpha_1^{\rm cy} \left(\frac{R}{M}\right)^2 + \alpha_2^{\rm cy} \left(\frac{R}{M}\right) - \alpha_3^{\rm cy},   
    \label{eq:dRcy-MR}
\end{equation}
where $\alpha_1^{\rm cy}$, $\alpha_2^{\rm cy}$, and $\alpha_3^{\rm cy}$ are positive
dimensionless adjustable coefficients that depend on $(L, K_0)$. 
We notice that $\Delta R_{\rm cy}$ reduces to zero in the case 
of $(K_0,-y)=(360\ {\rm MeV}, 220\ {\rm MeV\ fm}^3)$ in which
the spherical nuclei melt into uniform matter instead of changing
into cylindrical ones.   We also find that $\Delta R_{\rm cy}/R$ depends on 
the EOS parameters as strongly as $R/M$, which is a contrast to 
the case of $\Delta R_{\rm sp}/R$ but expected from the arguments based
on Eq.\ (\ref{eq:munexp}).

\begin{figure*}
\begin{center}
\begin{tabular}{ccc}
\includegraphics[scale=0.4]{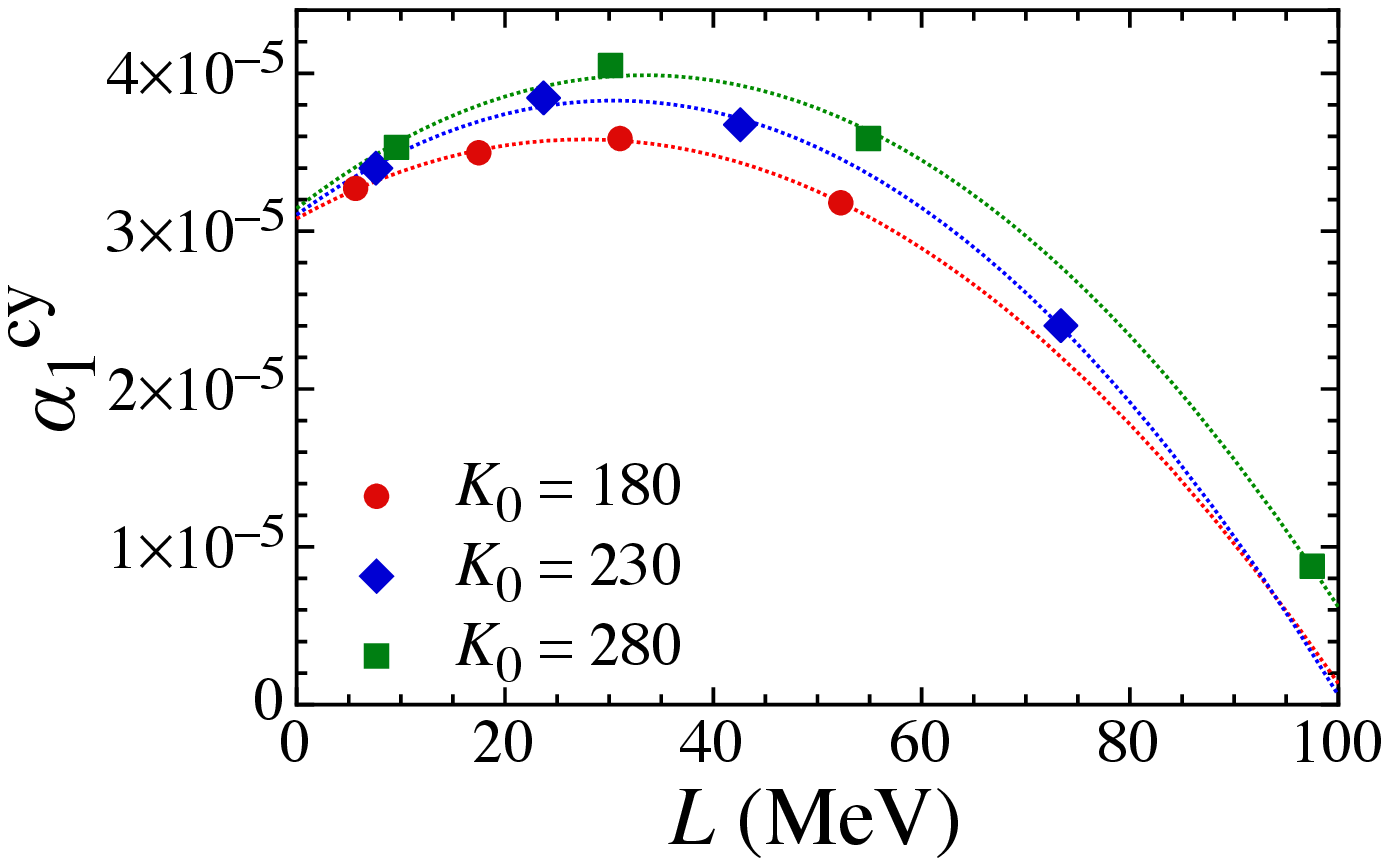} &
\includegraphics[scale=0.4]{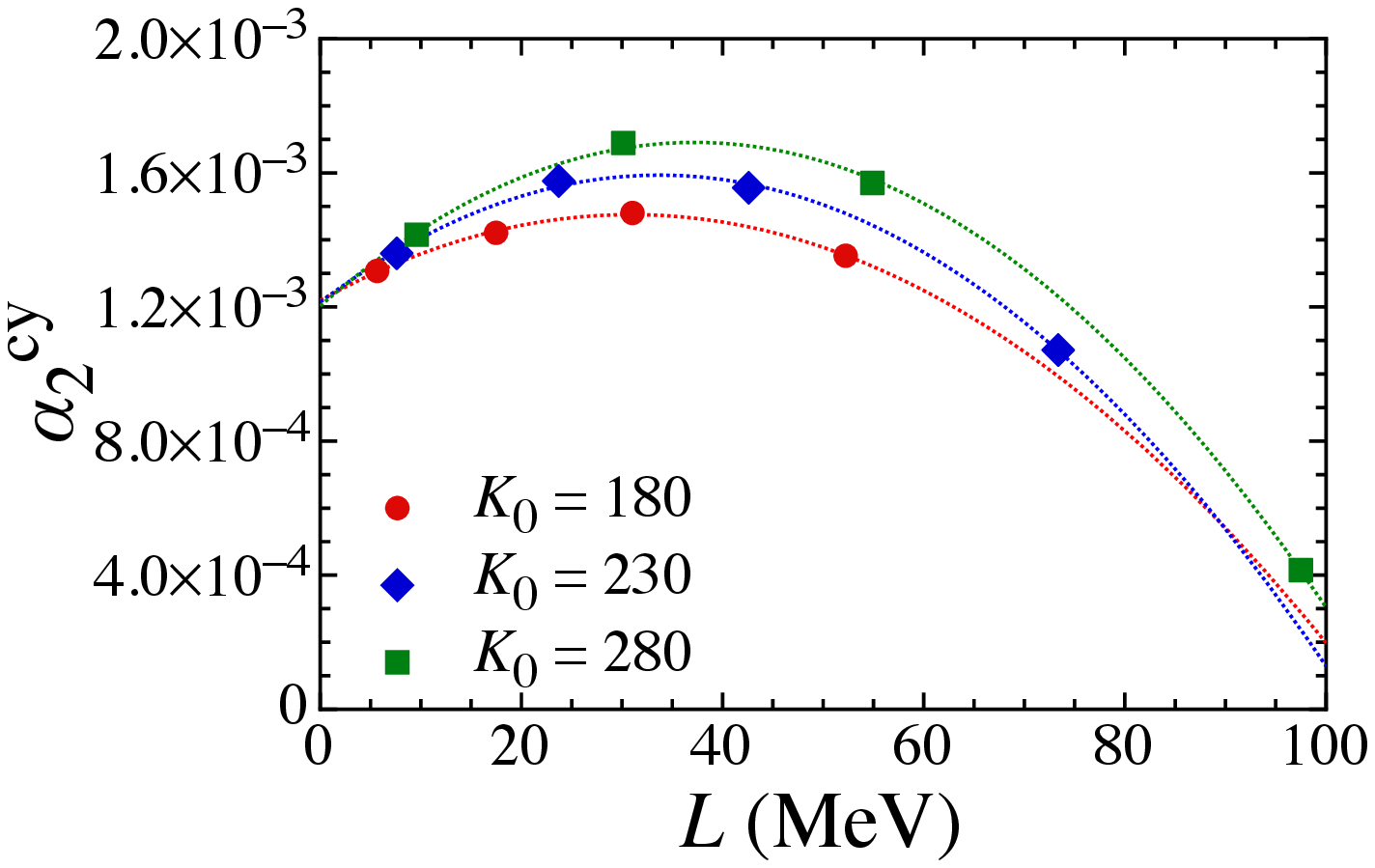} &
\includegraphics[scale=0.4]{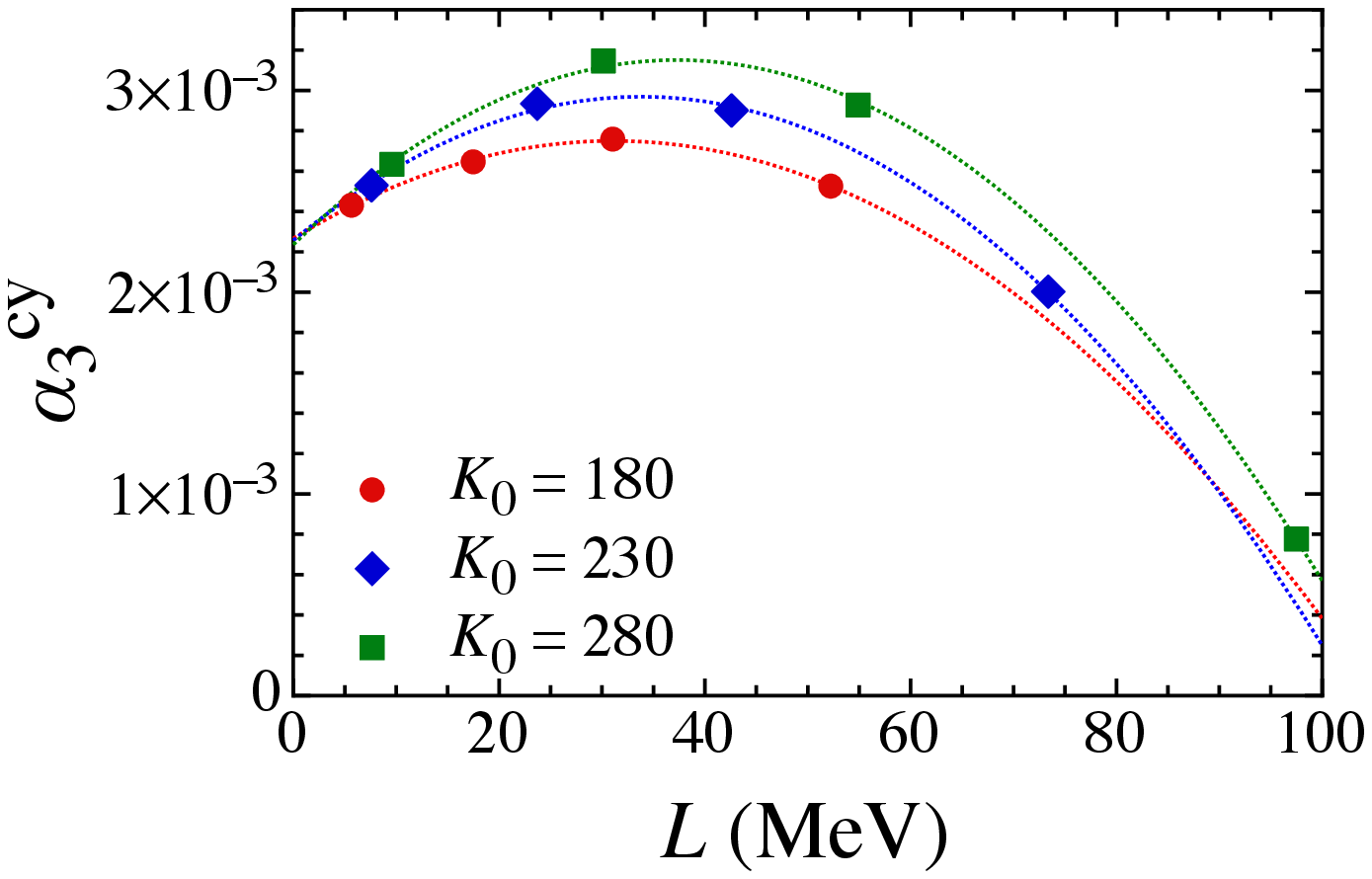} 
\end{tabular}
\end{center}
\caption{
Same as Fig.\ \ref{fig:asp-L}, but for the coefficients in Eq.\ 
(\ref{eq:dRcy-MR}) and the fitting formula given by Eqs.\ 
(\ref{eq:dRcy-a1})--(\ref{eq:dRcy-a3}).
}
\label{fig:acy-L}
\end{figure*}

In a similar way to the case of $\Delta R_{\rm sp}/R$, we plot in 
Fig.\ \ref{fig:acy-L} the values of $\alpha_1^{\rm cy}$, $\alpha_2^{\rm cy}$, and 
$\alpha_3^{\rm cy}$, which were obtained by fitting for several sets of 
$K_0$ and $L$, as a function of $L$.
These values can then be accurately expressed by the following fitting formulas:
\begin{eqnarray}
  \alpha_1^{\rm cy} &=& \beta_{11}^{\rm cy} + \beta_{12}^{\rm cy}\left(\frac{L}{60\ {\rm MeV}}\right) - \beta_{13}^{\rm cy}\left(\frac{L}{60\ {\rm MeV}}\right)^2,
       \label{eq:dRcy-a1} \\
  \alpha_2^{\rm cy} &=& \beta_{21}^{\rm cy} + \beta_{22}^{\rm cy}\left(\frac{L}{60\ {\rm MeV}}\right) - \beta_{23}^{\rm cy}\left(\frac{L}{60\ {\rm MeV}}\right)^2,
       \label{eq:dRcy-a2} \\  
  \alpha_3^{\rm cy} &=& \beta_{31}^{\rm cy} + \beta_{32}^{\rm cy}\left(\frac{L}{60\ {\rm MeV}}\right) - \beta_{33}^{\rm cy}\left(\frac{L}{60\ {\rm MeV}}\right)^2, 
       \label{eq:dRcy-a3}  
\end{eqnarray}
where $\beta_{ij}^{\rm cy}$ with $i=1$, 2, 3 and $j=1$, 2, 3 are positive 
dimensionless adjustable coefficients that depend on $K_0$. Again, 
such fitting does not work well for $K_0=360$ MeV.  We note that the 
functional form of $\alpha_{i}^{\rm cy}$ with respect to $L$ is different 
from that of $\alpha_{i}^{\rm sp}$.

\begin{figure}
\begin{center}
\includegraphics[scale=0.5]{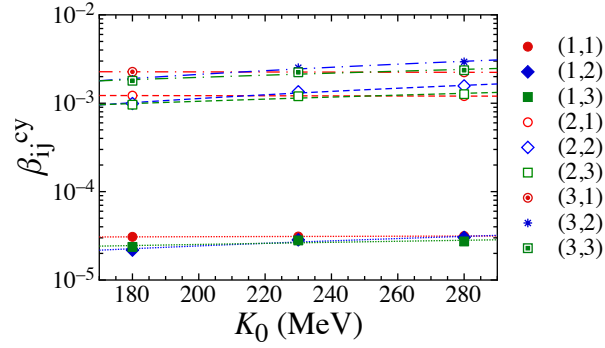}
\end{center}
\caption{
Same as Fig.\ \ref{fig:bsp-K0}, but for the coefficients in 
Eqs.\ (\ref{eq:dRcy-a1})--(\ref{eq:dRcy-a3}) and the corresponding 
fitting formulas.
}
\label{fig:bcy-K0}
\end{figure}

In Fig.\ \ref{fig:bcy-K0}, the values of $\beta_{ij}^{\rm cy}$ with 
$i=1$, 2, 3 and $j=1$, 2, 3, obtained by fitting for $K_0=180, 230, 280$ 
MeV, are shown as a function of $K_0$. From this figure, we find that 
$\beta_{ij}^{\rm cy}$ can be accurately expressed as a linear function 
of $K_0$,
\begin{eqnarray}
  \beta_{11}^{\rm cy} &=& \left[ 2.9604   +   0.1490\left(\frac{K_0}{230\ {\rm MeV}}\right)\right]\times 10^{-5},  \label{eq:dRcy-b11} \\
  \beta_{12}^{\rm cy} &=& \left[ 0.7194   +  1.9688\left(\frac{K_0}{230\ {\rm MeV}}\right)\right]\times 10^{-5},  \label{eq:dRcy-b12} \\
  \beta_{13}^{\rm cy} &=& \left[ 1.8003   +  0.8342\left(\frac{K_0}{230\ {\rm MeV}}\right)\right]\times 10^{-5},  \label{eq:dRcy-b13} \\
  \beta_{21}^{\rm cy} &=& \left[ 1.2506   - 0.03795\left(\frac{K_0}{230\ {\rm MeV}}\right)\right]\times 10^{-3},  \label{eq:dRcy-b21} \\
  \beta_{22}^{\rm cy} &=& \left[  -0.02827 + 1.3325\left(\frac{K_0}{230\ {\rm MeV}}\right)\right]\times 10^{-3},  \label{eq:dRcy-b22} \\
  \beta_{23}^{\rm cy} &=& \left[ 0.4456   +  0.6976\left(\frac{K_0}{230\ {\rm MeV}}\right)\right]\times 10^{-3},  \label{eq:dRcy-b23} \\
  \beta_{31}^{\rm cy} &=& \left[ 2.3235  - 0.06992\left(\frac{K_0}{230\ {\rm MeV}}\right)\right]\times 10^{-3},  \label{eq:dRcy-b31} \\
  \beta_{32}^{\rm cy} &=& \left[ -0.03548  +   2.4785\left(\frac{K_0}{230\ {\rm MeV}}\right)\right]\times 10^{-3},  \label{eq:dRcy-b32} \\
  \beta_{33}^{\rm cy} &=& \left[ 0.8266    +  1.3062\left(\frac{K_0}{230\ {\rm MeV}}\right)\right]\times 10^{-3}.  \label{eq:dRcy-b33} 
\end{eqnarray}
Now, we obtain a complete set of the fitting formulas 
(\ref{eq:dRcy-MR})--(\ref{eq:dRcy-b33}), which well reproduce 
the calculated values of $\Delta R_{\rm cy}/R$ for various
combinations of $R/M$, $L$, and $K_0$.  Note that 
applicability of these formulas is here again limited to 
the range of $180\lsim K_0\lsim 280$ MeV.

\subsection{Phases of slab-like (S), cylindrical-hole (CH), and spherical-hole 
(SH) nuclei}
\label{sec:III-C}

\begin{figure*}
\begin{center}
\begin{tabular}{cc}
\includegraphics[scale=0.5]{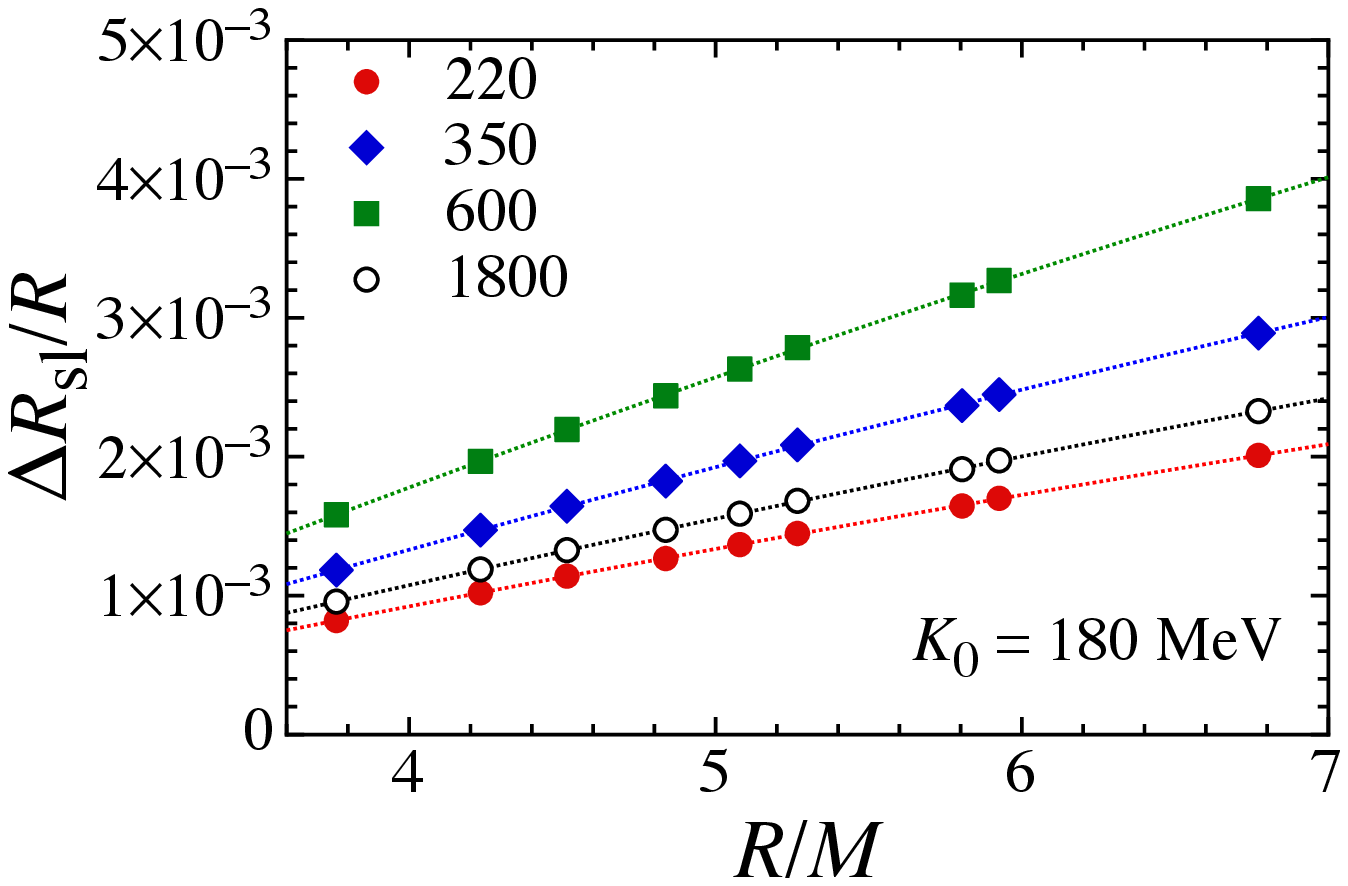} &
\includegraphics[scale=0.5]{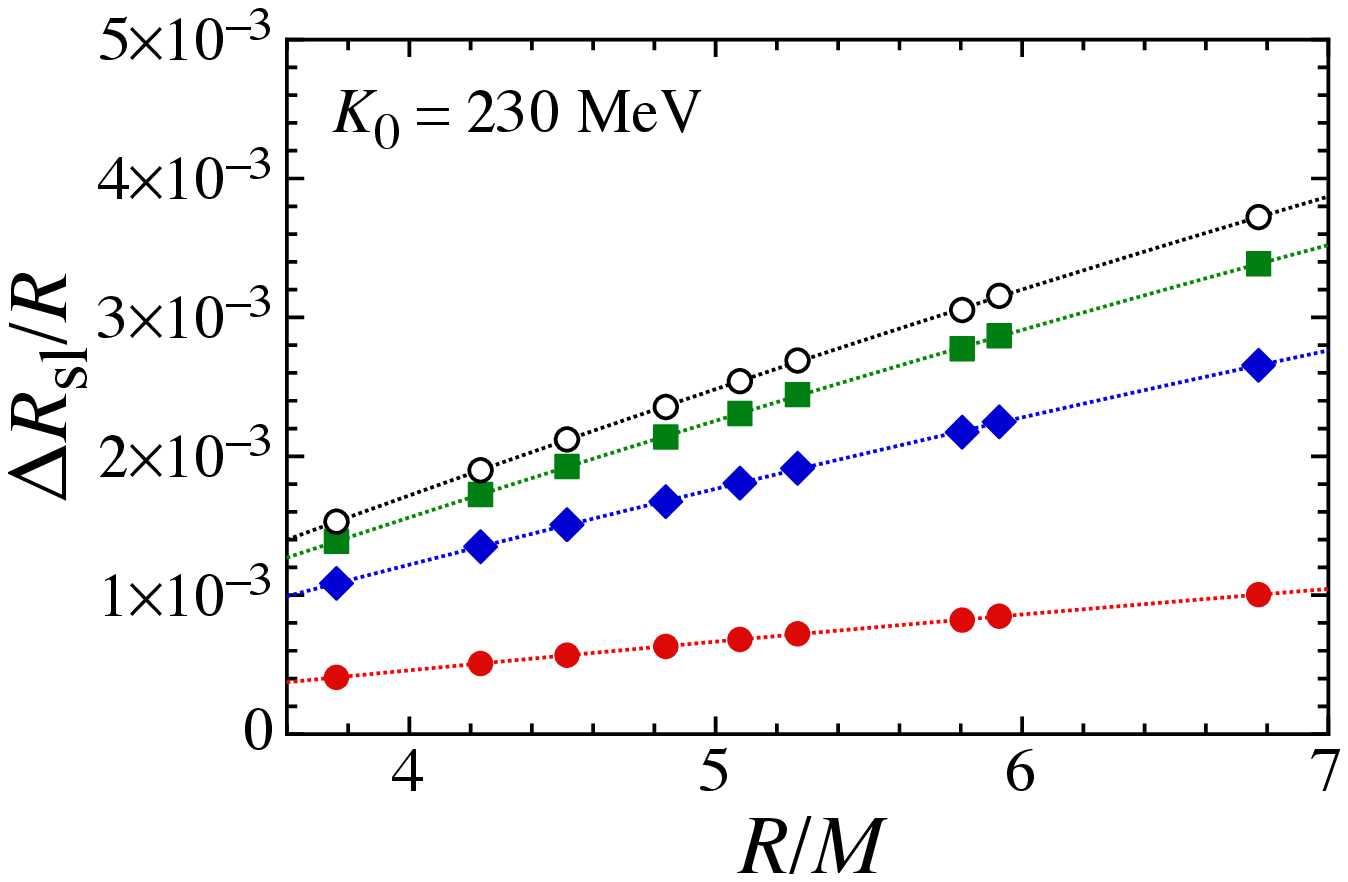} \\
\includegraphics[scale=0.5]{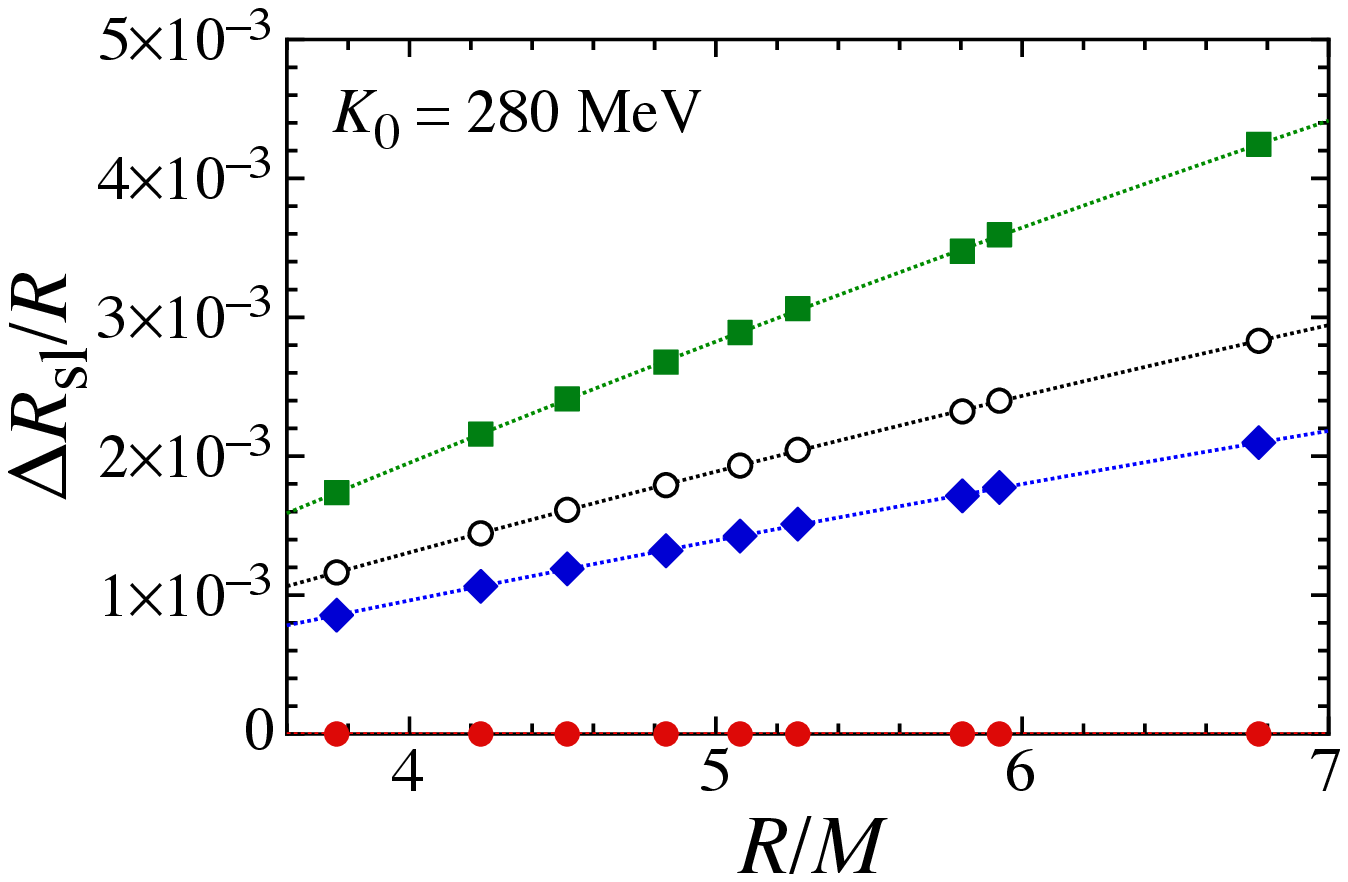} &
\includegraphics[scale=0.5]{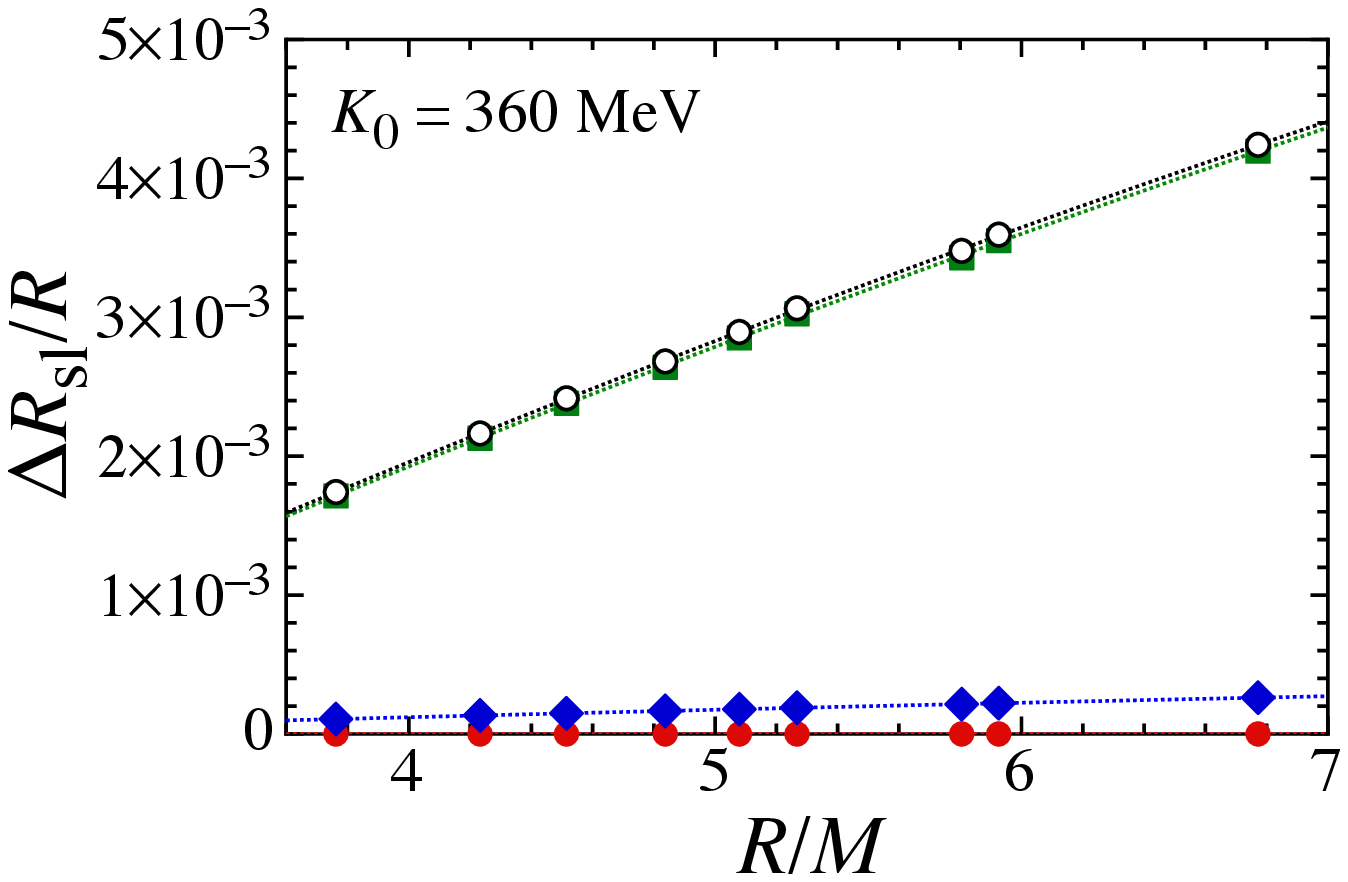} 
\end{tabular}
\end{center}
\caption{
Same as Fig.\ \ref{fig:dRsp-MR}, but for the ratio of the thickness of the 
phase composed of slab-like nuclei to the star's radius.  The dotted 
lines denote the fitting formula given by Eq.\ (\ref{eq:dRsl-MR}).
}
\label{fig:dRsl-MR}
\end{figure*}

\begin{figure*}
\begin{center}
\begin{tabular}{cc}
\includegraphics[scale=0.5]{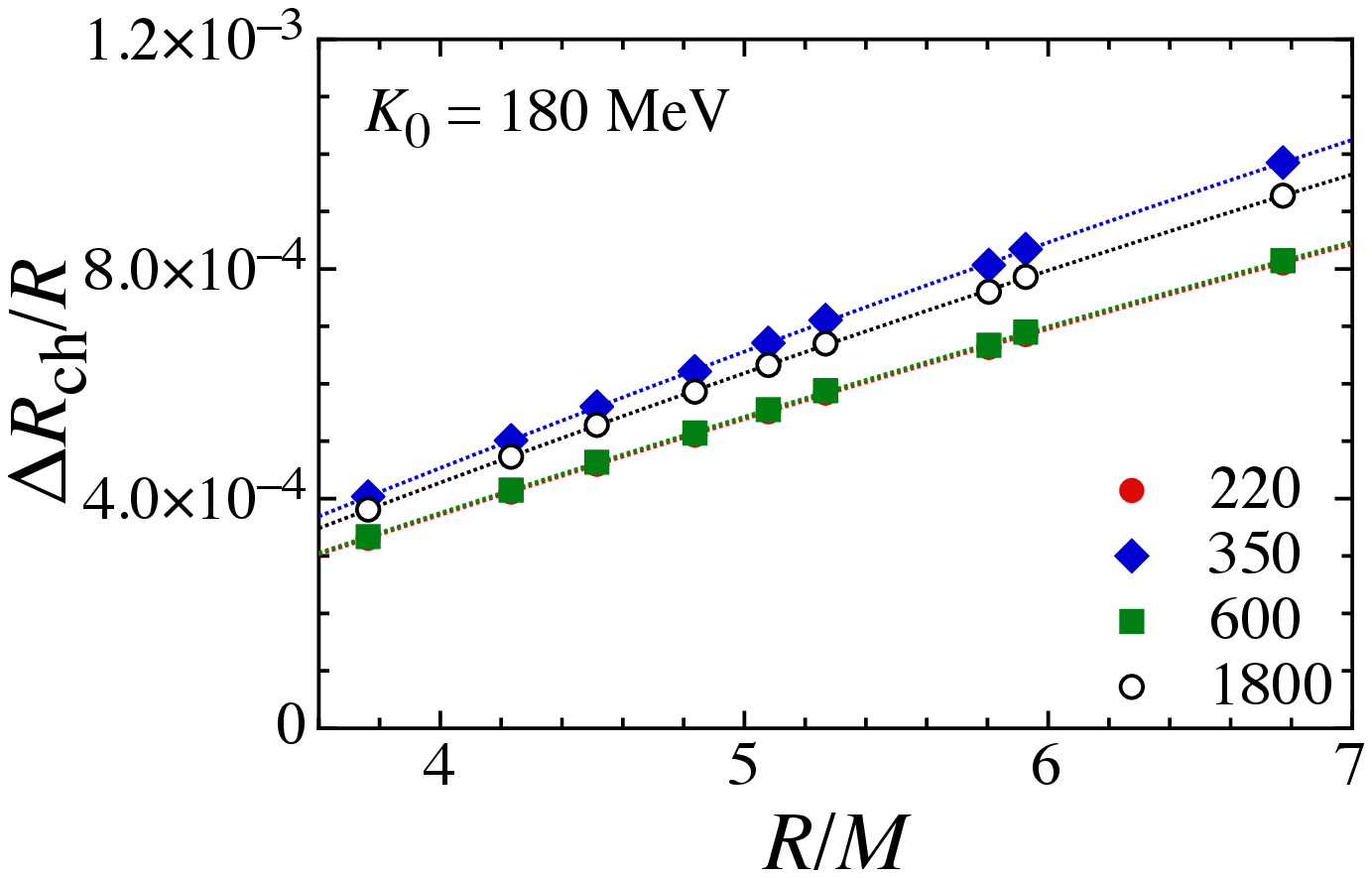} &
\includegraphics[scale=0.5]{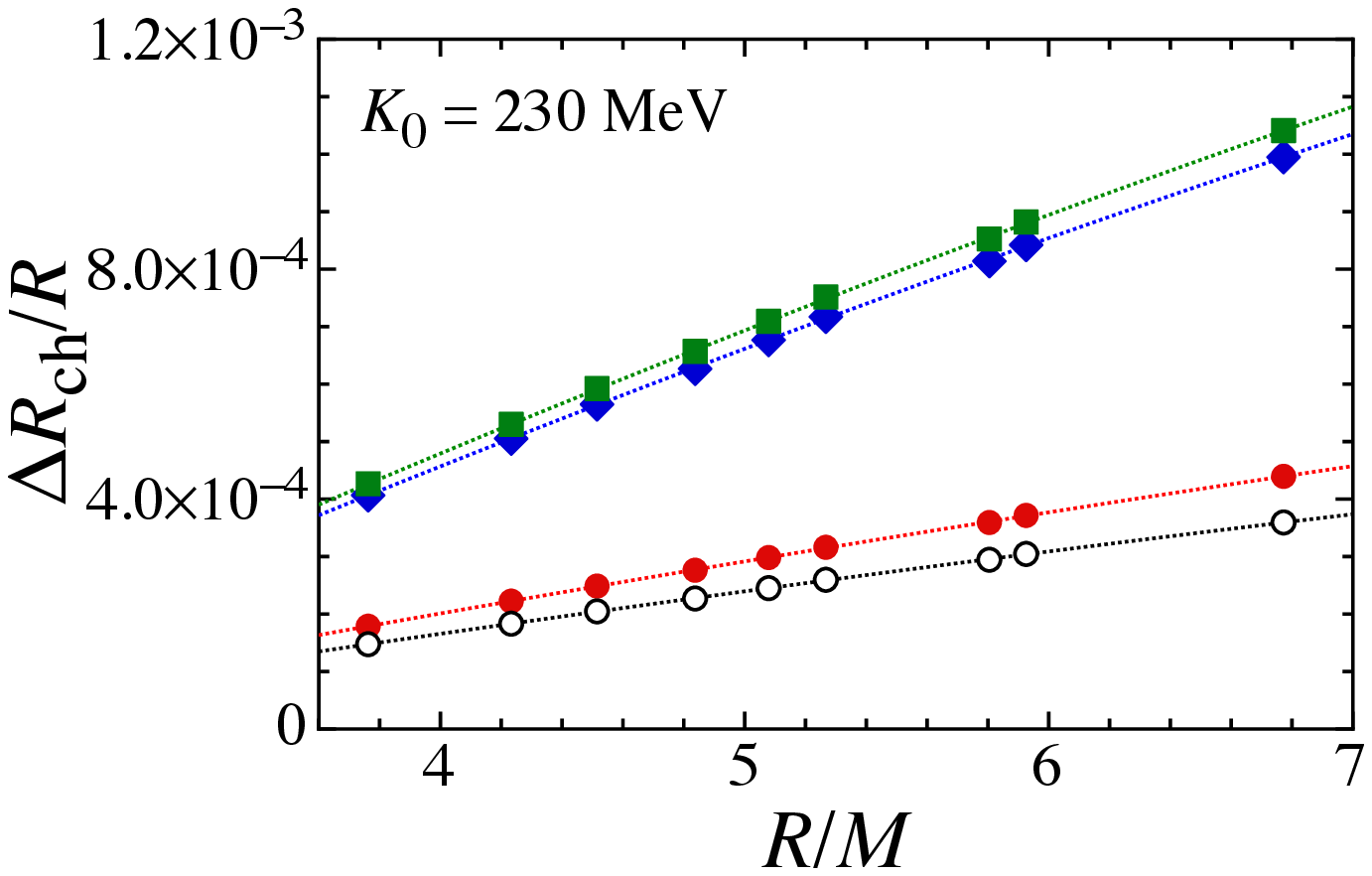} \\
\includegraphics[scale=0.5]{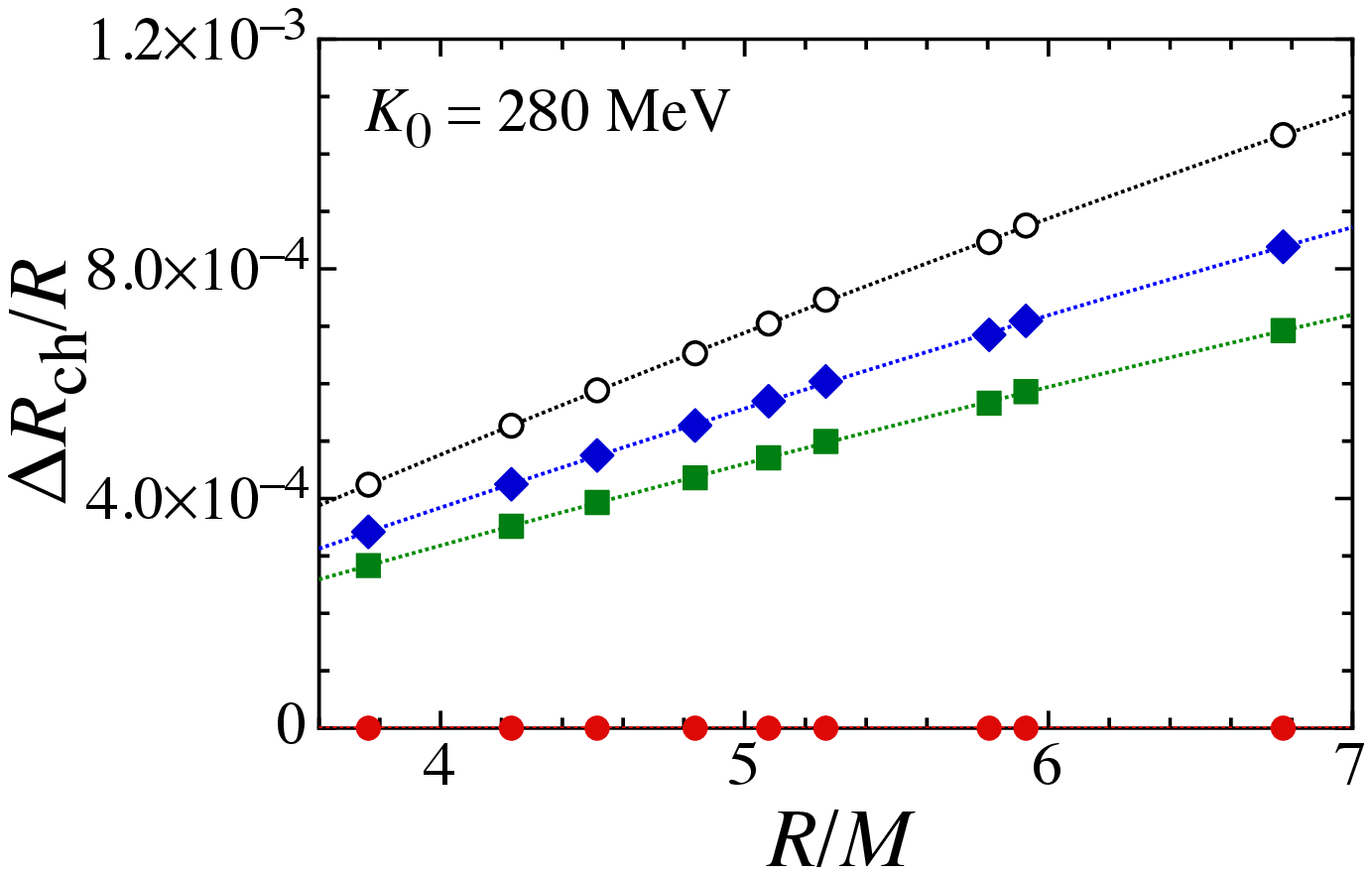} &
\includegraphics[scale=0.5]{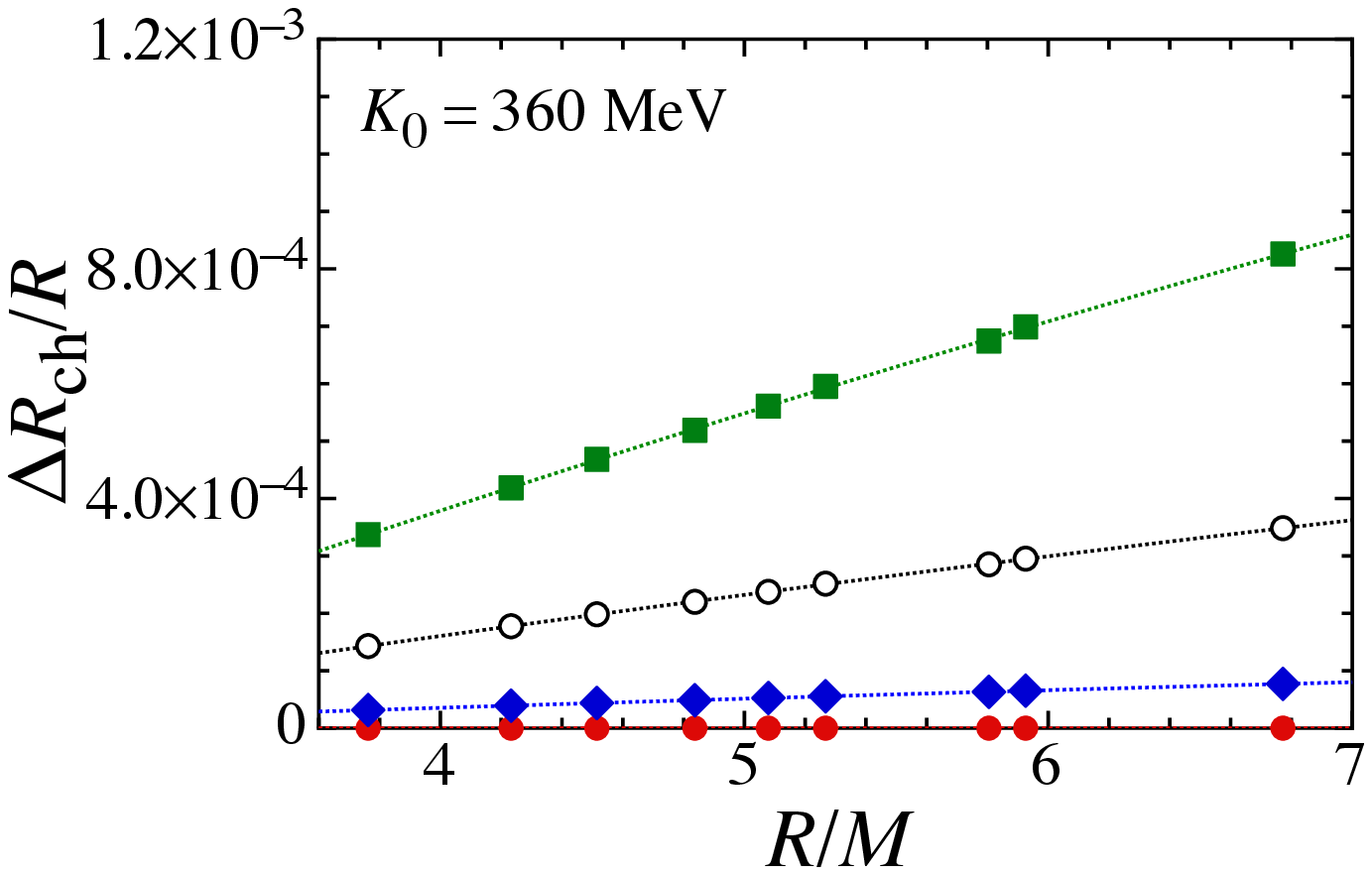} 
\end{tabular}
\end{center}
\caption{
Same as Fig.\ \ref{fig:dRsp-MR}, but for the ratio of the thickness of the 
phase composed of cylindrical-hole nuclei to the star's radius.  The 
dotted lines denote the fitting formula given by Eq.\ (\ref{eq:dRch-MR}).
}
\label{fig:dRch-MR}
\end{figure*}

\begin{figure*}
\begin{center}
\begin{tabular}{cc}
\includegraphics[scale=0.5]{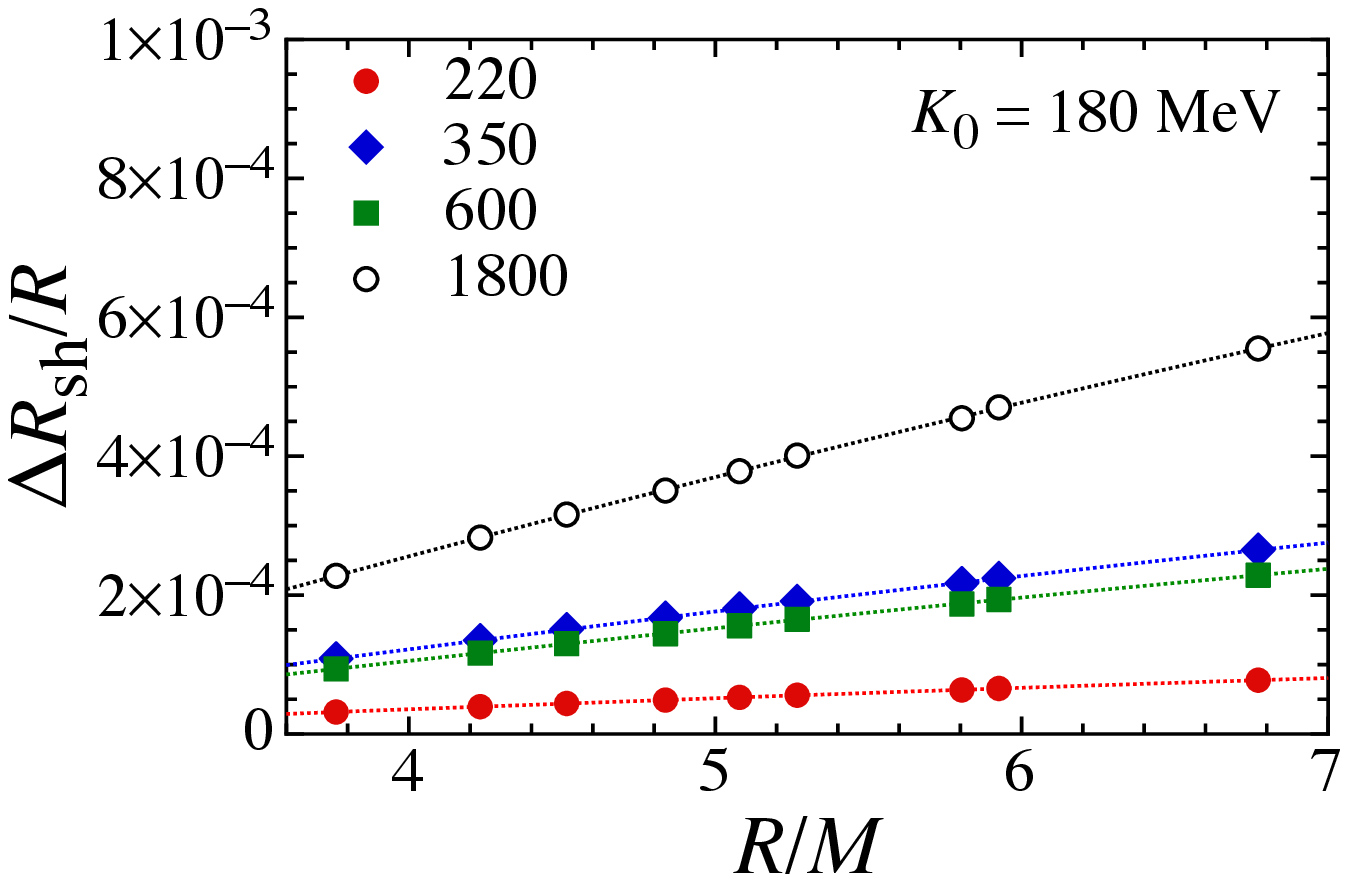} &
\includegraphics[scale=0.5]{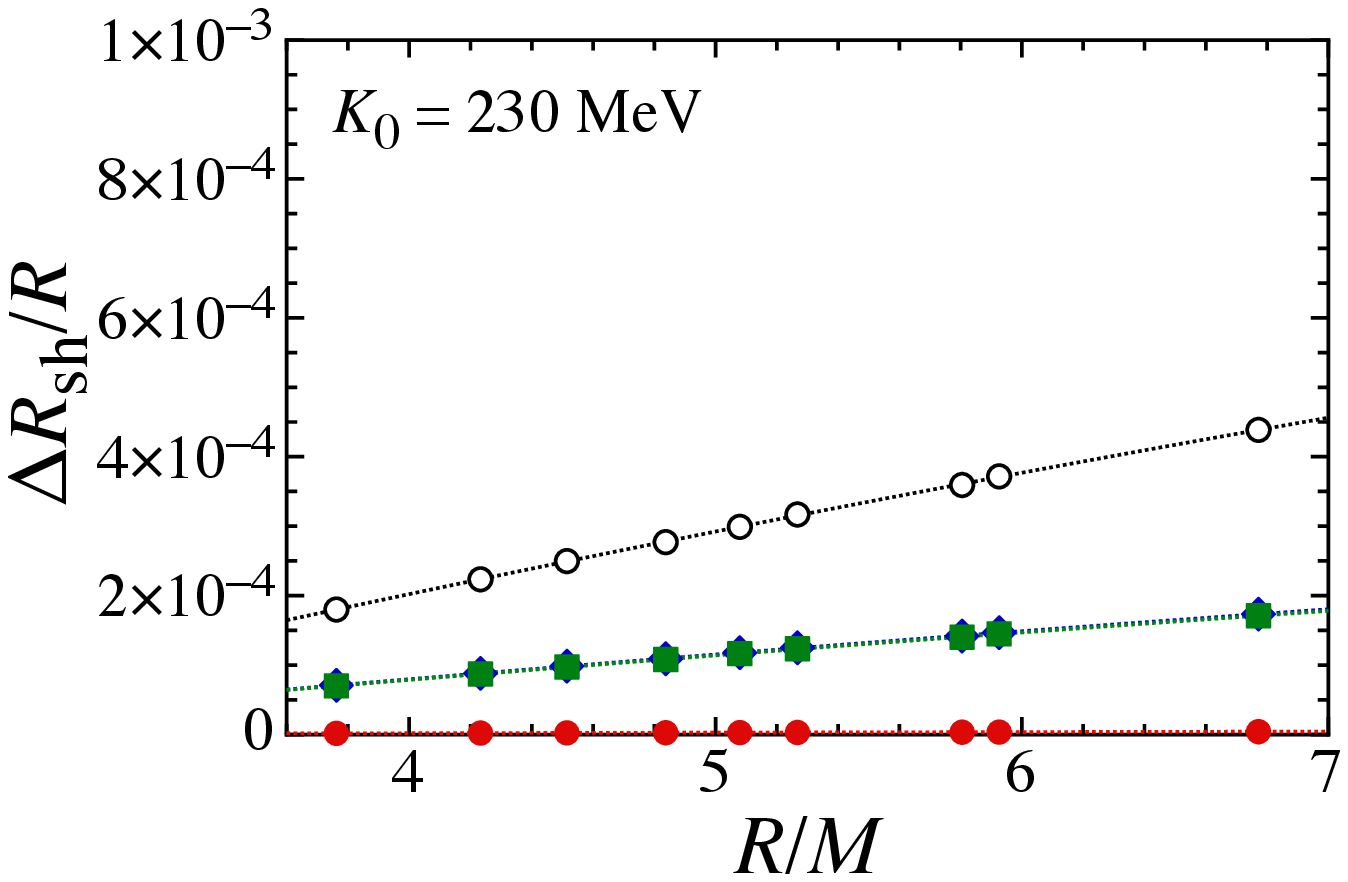} \\
\includegraphics[scale=0.5]{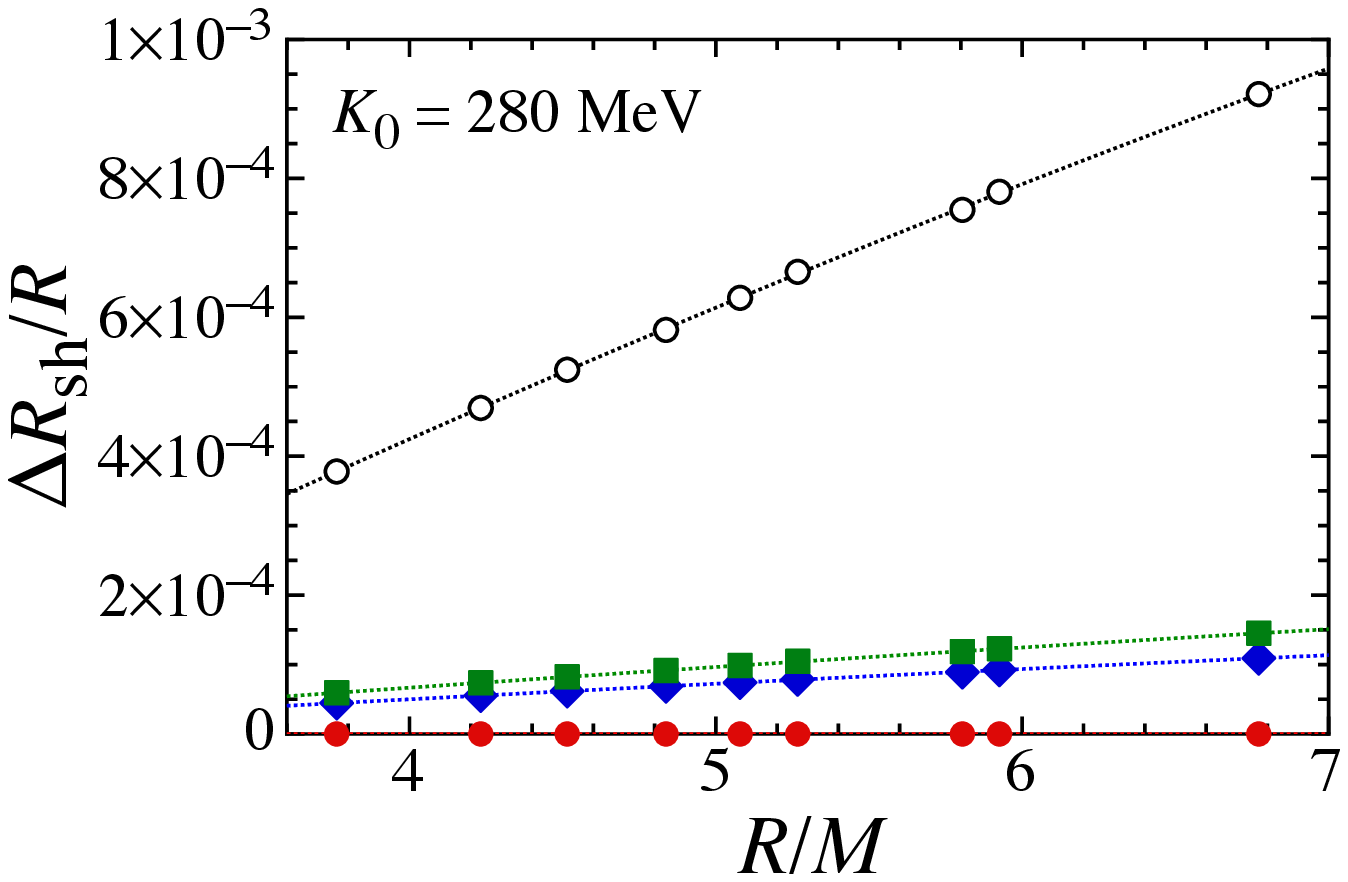} &
\includegraphics[scale=0.5]{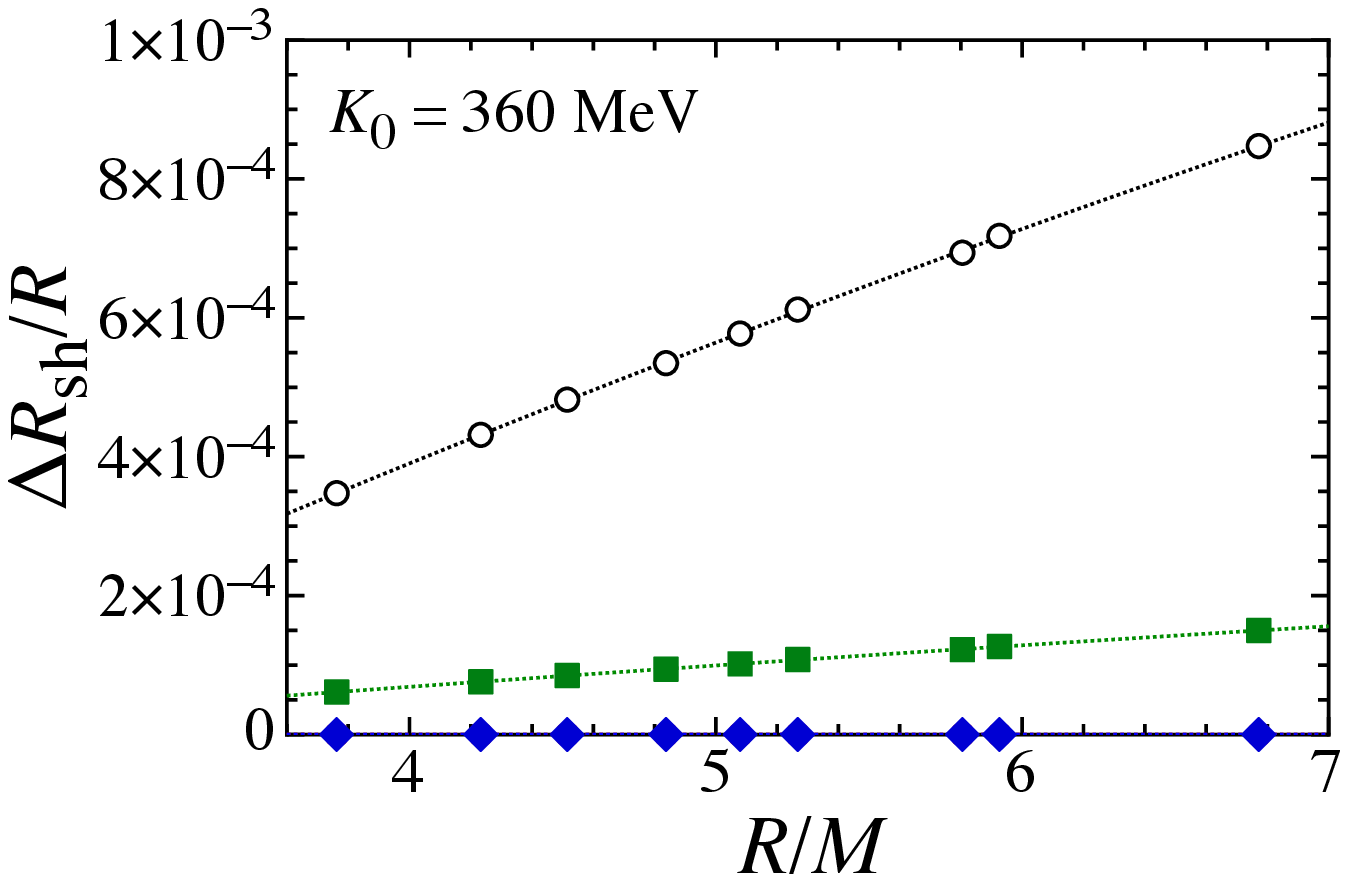} 
\end{tabular}
\end{center}
\caption{
Same as Fig.\ \ref{fig:dRsp-MR}, but for the ratio of the thickness of the 
phase composed of spherical-hole nuclei to the star's radius.  The 
dotted lines denote the fitting formula given by Eq.\ (\ref{eq:dRsh-MR}).
}
\label{fig:dRsh-MR}
\end{figure*}

We now focus on the rest of the pasta phases, namely, the S, CH, and SH
phases.  For various sets of the EOS parameters, the calculated relative
thickness of the layer of slab-like nuclei, $\Delta R_{\rm sl}/R$, 
that of the layer of cylindrical-hole nuclei, $\Delta R_{\rm ch}/R$, 
and that of the layer of spherical-hole nuclei, $\Delta R_{\rm sh}/R$,
are shown in Figs.\ \ref{fig:dRsl-MR}, \ref{fig:dRch-MR}, and \ref{fig:dRsh-MR},
respectively, as a function of $R/M$.  From these figures, we can confirm
that $\Delta R_{\rm sl}/R$, $\Delta R_{\rm ch}/R$, and $\Delta R_{\rm sh}/R$ can 
be accurately expressed as a function of $R/M$ by
\begin{eqnarray}
  \frac{\Delta R_{\rm sl}}{R} &=& -\alpha_1^{\rm sl} \left(\frac{R}{M}\right)^2 + \alpha_2^{\rm sl} \left(\frac{R}{M}\right) - \alpha_3^{\rm sl},
     \label{eq:dRsl-MR} \\
  \frac{\Delta R_{\rm ch}}{R} &=& -\alpha_1^{\rm ch} \left(\frac{R}{M}\right)^2 + \alpha_2^{\rm ch} \left(\frac{R}{M}\right) - \alpha_3^{\rm ch},
     \label{eq:dRch-MR} \\
  \frac{\Delta R_{\rm sh}}{R} &=& -\alpha_1^{\rm sh} \left(\frac{R}{M}\right)^2 + \alpha_2^{\rm sh} \left(\frac{R}{M}\right) - \alpha_3^{\rm sh},
    \label{eq:dRsh-MR}
\end{eqnarray}
where $\alpha_i^{\rm sl}$, $\alpha_i^{\rm ch}$, and $\alpha_i^{\rm sh}$ with $i=1$, 
2, 3 are positive dimensionless adjustable coefficients, tabulated
in Tables \ref{tab:asl}, \ref{tab:ach}, and \ref{tab:ash}.  One can observe that $\Delta R_{\rm sl}/R$, 
$\Delta R_{\rm ch}/R$, and $\Delta R_{\rm sh}/R$ depend on the EOS parameters
as strongly as $R/M$, as in the case of $\Delta R_{\rm cy}/R$. 
We remark that in contrast to the cases of $\Delta R_{\rm sp}/R$ 
and $\Delta R_{\rm cy}/R$, the coefficients in Eqs.\ 
(\ref{eq:dRsl-MR})--(\ref{eq:dRsh-MR}) are difficult to express as a simple 
function of $(L, K_0)$.  This is mainly because the thickness of each 
layer, which is tiny or even zero, has a complicated dependence on 
the EOS parameters.

\begin{table}
\centering
\begin{minipage}{100mm}
\caption{
The optimized coefficients in Eq.\ (\ref{eq:dRsl-MR}).
}
\begin{tabular}{ccccc}
\hline\hline
  $K_0$ (MeV) & $L$ (MeV) 
     & $\alpha_1^{\rm sl}$ & $\alpha_s^{\rm sl}$ & $\alpha_3^{\rm sl}$ \\
\hline
  180 & 5.7 & $1.502\times 10^{-5}$ & $6.138\times 10^{-4}$ & $1.140\times 10^{-3}$      \\  
  180 & 17.5   & $2.432\times 10^{-5}$ & $1.012\times 10^{-3}$ & $1.881\times 10^{-3}$      \\  
  180 & 31.0   & $1.798\times 10^{-5}$ & $7.561\times 10^{-4}$ & $1.406\times 10^{-3}$      \\  
  180 & 52.2   & $1.211\times 10^{-5}$ & $5.226\times 10^{-4}$ & $9.740\times 10^{-4}$      \\  
\hline
  230 & 7.6 & $2.389\times 10^{-5}$ & $9.801\times 10^{-4}$ & $1.820\times 10^{-3}$       \\  
  230 & 23.7   & $2.113\times 10^{-5}$ & $8.862\times 10^{-4}$ & $1.647\times 10^{-3}$       \\  
  230 & 42.6   & $1.596\times 10^{-5}$ & $6.896\times 10^{-4}$ & $1.283\times 10^{-3}$       \\  
  230 & 73.4   & $5.723\times 10^{-6}$ & $2.582\times 10^{-4}$ & $4.819\times 10^{-4}$       \\  
\hline
  280 & 9.6 & $1.822\times 10^{-5}$ & $7.460\times 10^{-4}$ & $1.385\times 10^{-3}$       \\  
  280 & 30.1   & $2.568\times 10^{-5}$ & $1.104\times 10^{-3}$ & $2.051\times 10^{-3}$       \\  
  280 & 54.9   & $1.208\times 10^{-5}$ & $5.399\times 10^{-4}$ & $1.005\times 10^{-3}$       \\  
  280 & 97.5   & $0.0$ & $0.0$ & $0.0$       \\  
\hline
  360 & 12.8 & $2.691\times 10^{-5}$ & $1.114\times 10^{-3}$ & $2.068\times 10^{-3}$       \\  
  360 & 40.9   & $2.418\times 10^{-5}$ & $1.079\times 10^{-3}$ & $2.005\times 10^{-3}$       \\  
  360 & 76.4   & $1.660\times 10^{-6}$ & $6.874\times 10^{-5}$ & $1.279\times 10^{-4}$       \\  
  360 & 146.1   & $0.0$ & $0.0$ & $0.0$       \\  
\hline\hline
\end{tabular}
\label{tab:asl}
\end{minipage}
\end{table}

\begin{table}
\centering
\begin{minipage}{100mm}
\caption{
The optimized coefficients in Eq.\ (\ref{eq:dRch-MR}).
}
\begin{tabular}{ccccc}
\hline\hline
  $K_0$ (MeV) & $L$ (MeV)
     & $\alpha_1^{\rm ch}$ & $\alpha_s^{\rm ch}$ & $\alpha_3^{\rm ch}$ \\
\hline
  180 & 5.7 & $5.827\times 10^{-6}$ & $2.430\times 10^{-4}$ & $4.510\times 10^{-4}$      \\  
  180 & 17.5   & $5.002\times 10^{-6}$ & $2.123\times 10^{-4}$ & $3.942\times 10^{-4}$      \\  
  180 & 31.0   & $6.025\times 10^{-6}$ & $2.567\times 10^{-4}$ & $4.771\times 10^{-4}$      \\  
  180 & 52.2   & $4.837\times 10^{-6}$ & $2.102\times 10^{-4}$ & $3.916\times 10^{-4}$      \\  
\hline
  230 & 7.6 & $2.244\times 10^{-6}$ & $9.401\times 10^{-5}$ & $1.744\times 10^{-4}$       \\  
  230 & 23.7   & $6.347\times 10^{-6}$ & $2.710\times 10^{-4}$ & $5.032\times 10^{-4}$       \\  
  230 & 42.6   & $5.886\times 10^{-6}$ & $2.575\times 10^{-4}$ & $4.787\times 10^{-4}$       \\  
  230 & 73.4   & $2.494\times 10^{-6}$ & $1.129\times 10^{-4}$ & $2.107\times 10^{-4}$       \\  
\hline
  280 & 9.6 & $6.487\times 10^{-6}$ & $2.706\times 10^{-4}$ & $5.023\times 10^{-4}$       \\  
  280 & 30.1   & $4.079\times 10^{-6}$ & $1.789\times 10^{-4}$ & $3.321\times 10^{-4}$       \\  
  280 & 54.9   & $4.766\times 10^{-6}$ & $2.152\times 10^{-4}$ & $4.002\times 10^{-4}$       \\  
  280 & 97.5   & $0.0$ & $0.0$ & $0.0$       \\  
\hline
  360 & 12.8 & $2.146\times 10^{-6}$ & $9.086\times 10^{-5}$ & $1.685\times 10^{-4}$       \\  
  360 & 40.9   & $4.630\times 10^{-6}$ & $2.112\times 10^{-4}$ & $3.918\times 10^{-4}$       \\  
  360 & 76.4   & $4.882\times 10^{-7}$ & $2.024\times 10^{-5}$ & $3.763\times 10^{-5}$       \\  
  360 & 146.1   & $0.0$ & $0.0$ &$0.0$        \\  
\hline\hline
\end{tabular}
\label{tab:ach}
\end{minipage}
\end{table}

\begin{table}
\centering
\begin{minipage}{100mm}
\caption{
The optimized coefficients in Eq.\ (\ref{eq:dRsh-MR}).
}
\begin{tabular}{ccccc}
\hline\hline
  $K_0$ (MeV) & $L$ (MeV)
     & $\alpha_1^{\rm sh}$ & $\alpha_s^{\rm sh}$ & $\alpha_3^{\rm sh}$ \\
\hline
  180 & 5.7 & $3.435\times 10^{-6}$ & $1.449\times 10^{-4}$ & $2.689\times 10^{-4}$      \\  
  180 & 17.5   & $1.395\times 10^{-6}$ & $5.952\times 10^{-5}$ & $1.105\times 10^{-4}$      \\  
  180 & 31.0   & $1.610\times 10^{-6}$ & $6.892\times 10^{-5}$ & $1.281\times 10^{-4}$      \\  
  180 & 52.2   & $4.609\times 10^{-7}$ & $2.008\times 10^{-5}$ & $3.740\times 10^{-5}$      \\  
\hline
  230 & 7.6 & $2.718\times 10^{-6}$ & $1.146\times 10^{-4}$ & $2.125\times 10^{-4}$       \\  
  230 & 23.7   & $1.035\times 10^{-6}$ & $4.446\times 10^{-5}$ & $8.253\times 10^{-5}$       \\  
  230 & 42.6   & $1.020\times 10^{-6}$ & $4.479\times 10^{-5}$ & $8.326\times 10^{-5}$       \\  
  230 & 73.4   & $2.312\times 10^{-8}$ & $1.048\times 10^{-6}$ & $1.955\times 10^{-6}$       \\  
\hline
  280 & 9.6 & $5.688\times 10^{-6}$ & $2.405\times 10^{-4}$ & $4.462\times 10^{-4}$       \\  
  280 & 30.1   & $8.503\times 10^{-7}$ & $3.744\times 10^{-5}$ & $6.949\times 10^{-5}$       \\  
  280 & 54.9   & $6.170\times 10^{-7}$ & $2.795\times 10^{-5}$ & $5.198\times 10^{-5}$       \\  
  280 & 97.5   & $0.0$ & $0.0$ & $0.0$       \\  
\hline
  360 & 12.8 & $5.171\times 10^{-6}$ & $2.205\times 10^{-4}$ & $4.091\times 10^{-4}$       \\  
  360 & 40.9   & $8.349\times 10^{-7}$ & $3.825\times 10^{-5}$ & $7.096\times 10^{-5}$       \\  
  360 & 76.4   & $0.0$ & $0.0$ & $0.0$       \\  
  360 & 146.1   & $0.0$ & $0.0$ & $0.0$       \\  
\hline\hline
\end{tabular}
\label{tab:ash}
\end{minipage}
\end{table}

\subsection{Crust thickness}
\label{sec:III-D}

We conclude this section by deriving a simple fitting formula for the
ratio of the crust thickness to the star's radius.  Such derivation may well 
be possible even though $\Delta R_{\rm sl}/R$, $\Delta R_{\rm ch}/R$, and 
$\Delta R_{\rm sh}/R$ are difficult to express by a simple fitting 
formula.  This is because the crust thickness, $\Delta R$, is dominated
by the phase composed of spherical nuclei, of which the thickness has 
been parametrized above.

\begin{figure*}
\begin{center}
\begin{tabular}{cc}
\includegraphics[scale=0.5]{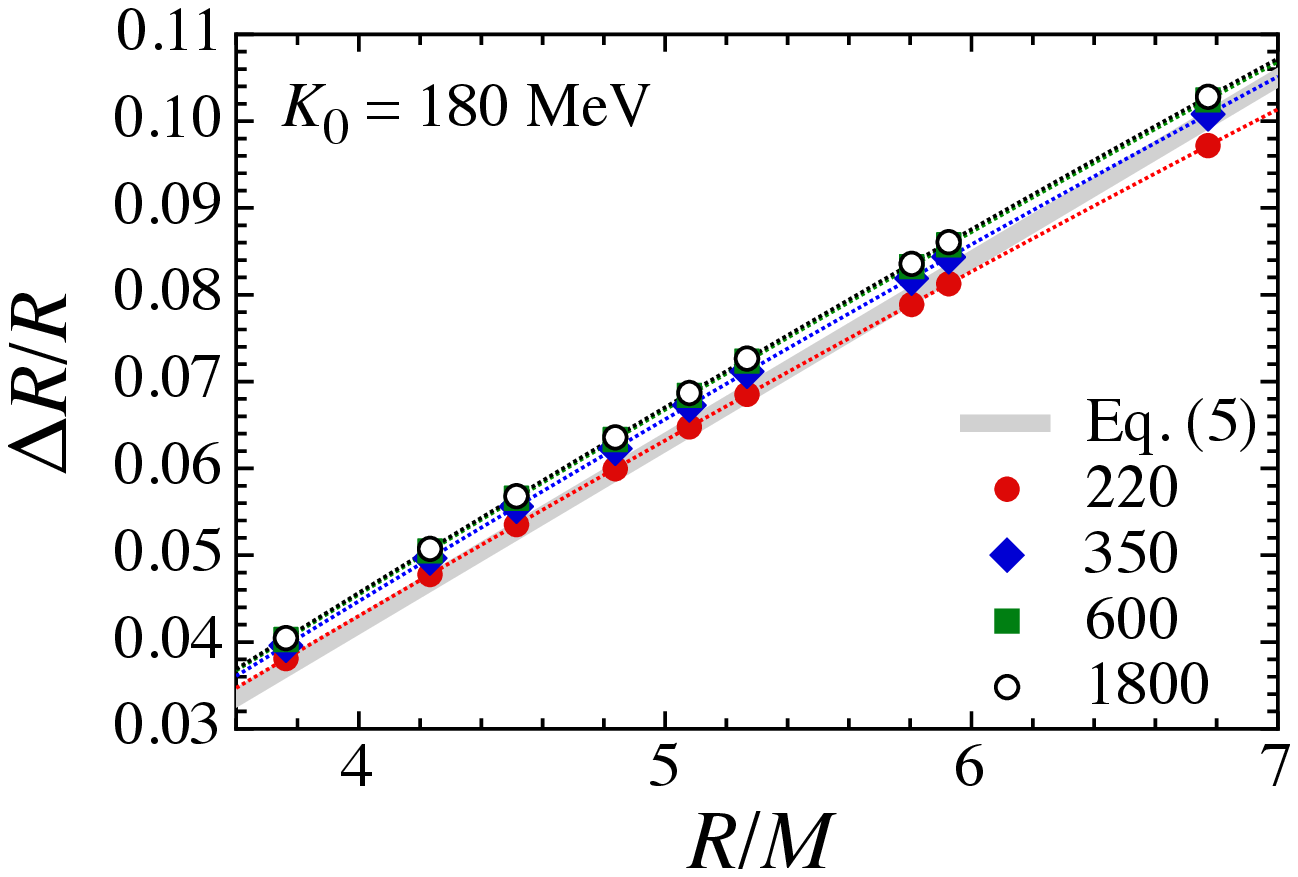} &
\includegraphics[scale=0.5]{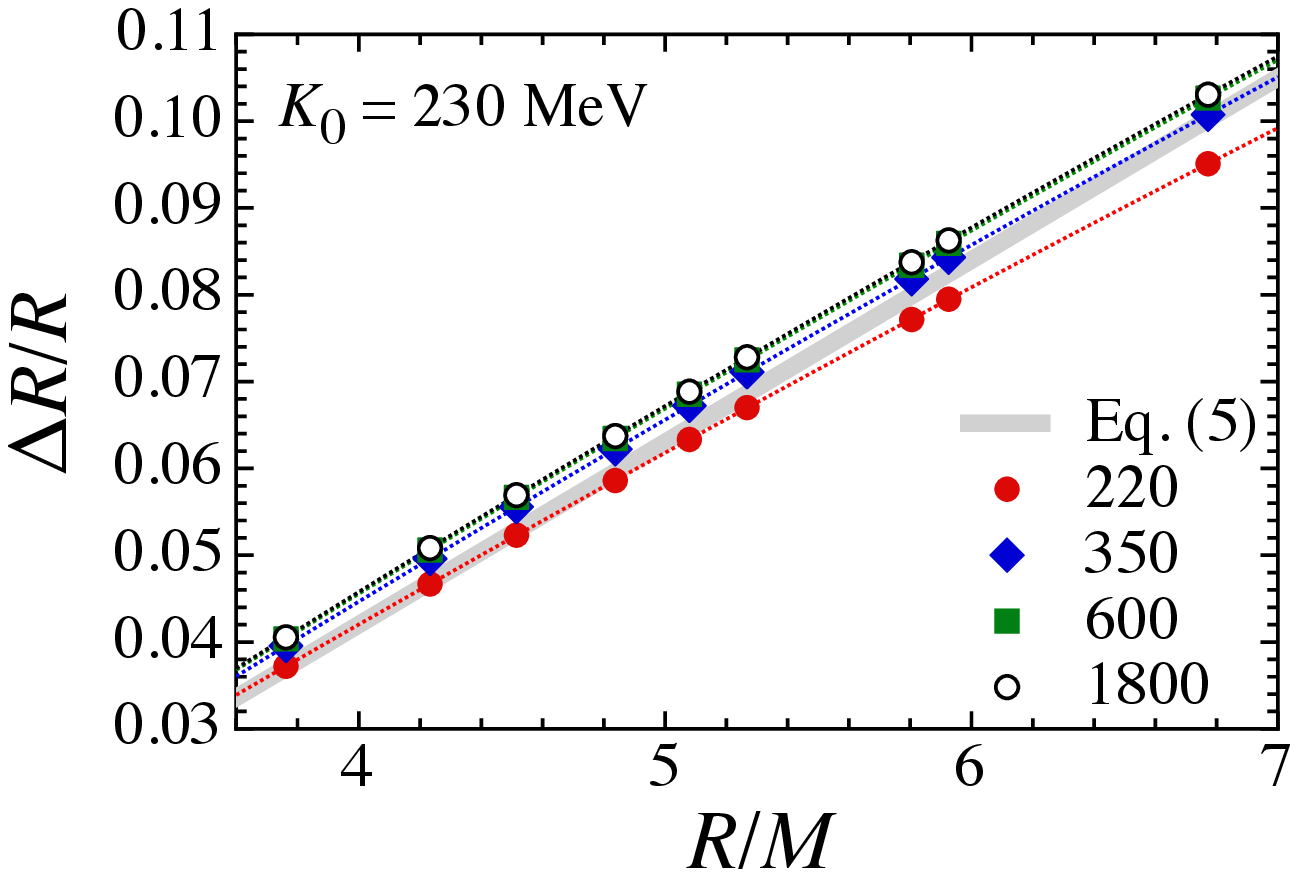} \\
\includegraphics[scale=0.5]{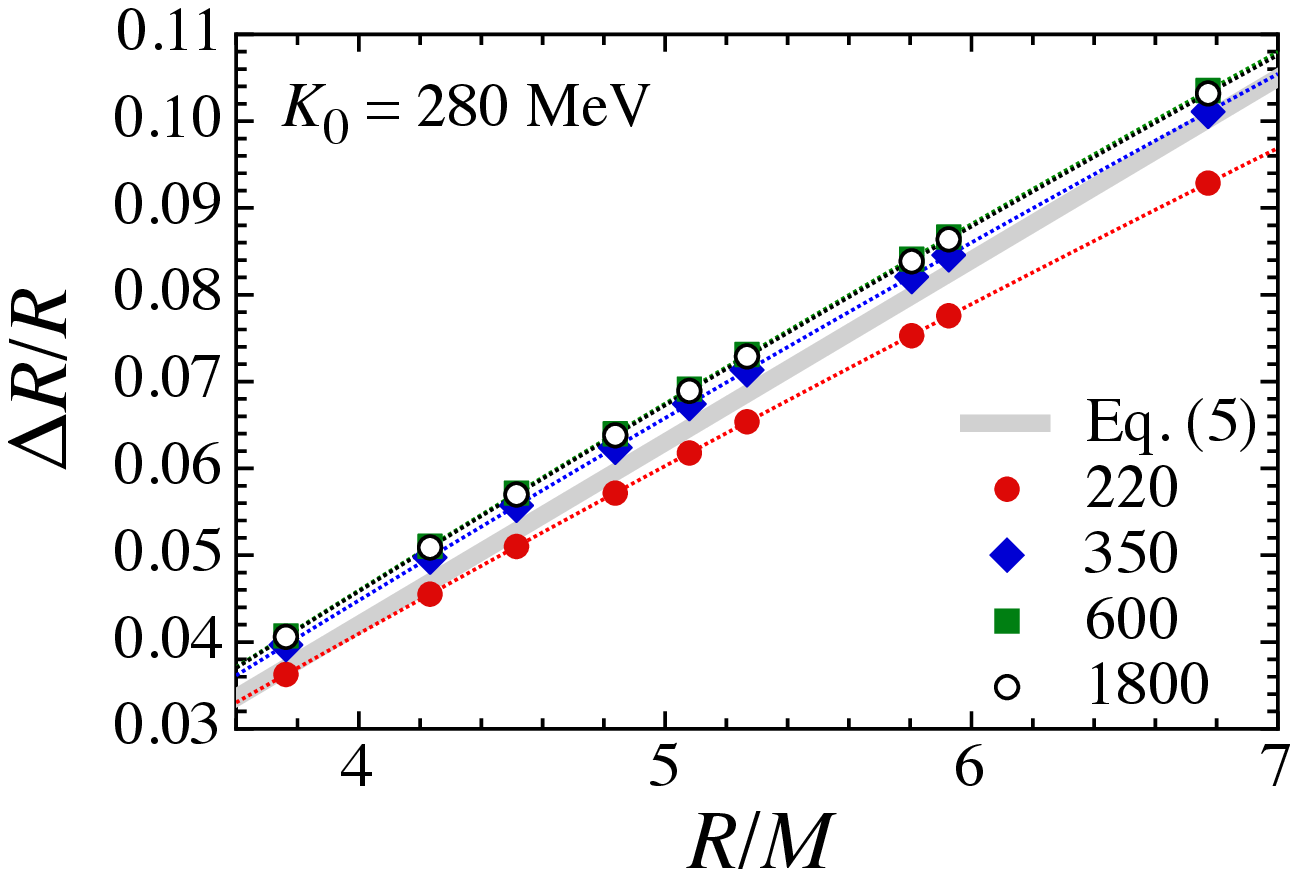} &
\includegraphics[scale=0.5]{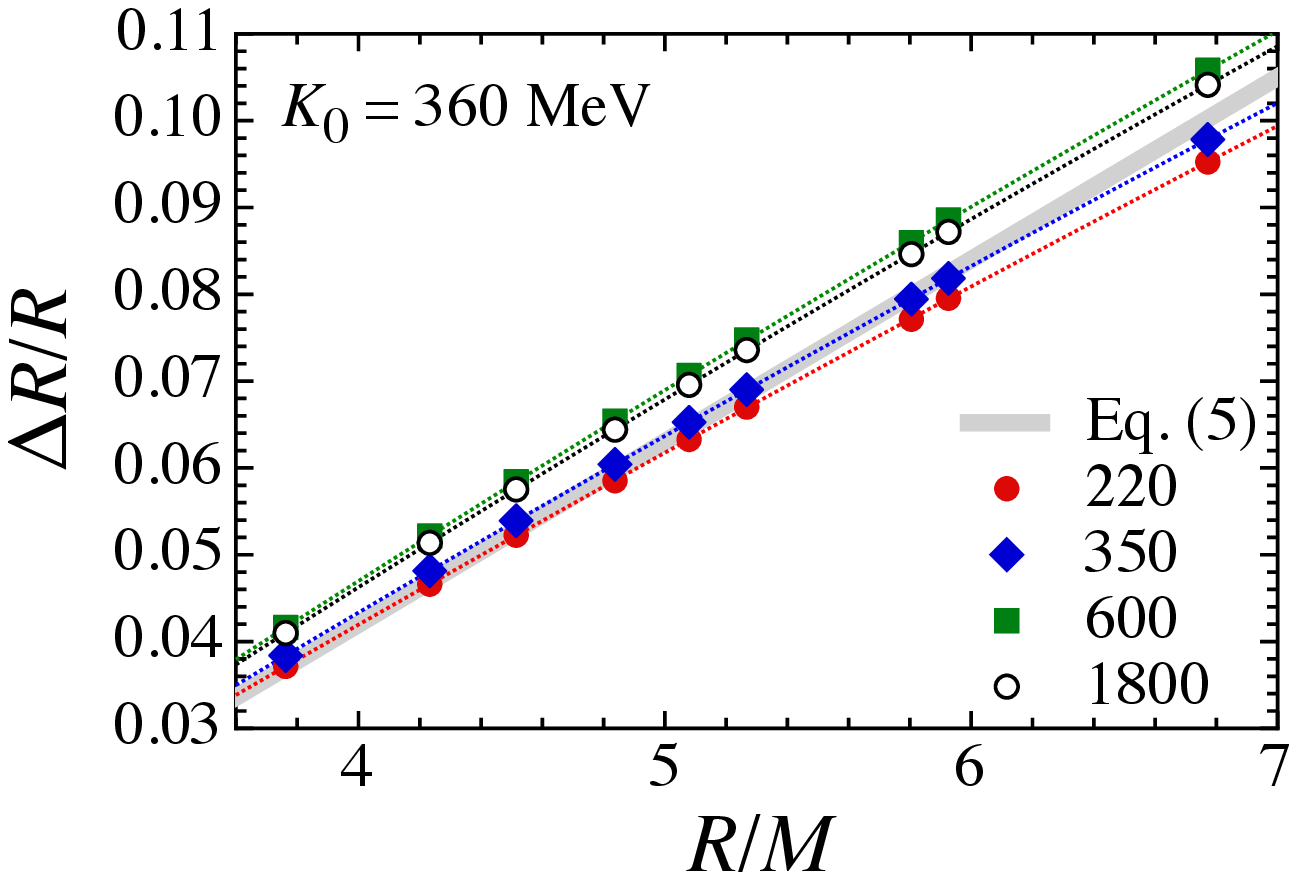} 
\end{tabular}
\end{center}
\caption{
Same as Fig.\ \ref{fig:dRsp-MR}, but for the ratio of the crust thickness 
to the star's radius.  The dotted lines denote the fitting formula 
given by Eq.\ (\ref{eq:dR-MR}), while the thick-solid line denotes the typical 
crust thickness given by Eq.\ (\ref{eq:typ}).
}
\label{fig:dR-MR}
\end{figure*}

In Fig.\ \ref{fig:dR-MR}, we plot the results for $\Delta R/R$ 
obtained for various EOS parameters as a function of $R/M$, which 
can be accurately expressed by
\begin{equation}
  \frac{\Delta R}{R} = -\alpha_1 \left(\frac{R}{M}\right)^2 + \alpha_2 \left(\frac{R}{M}\right) - \alpha_3,  \label{eq:dR-MR}
\end{equation}
where $\alpha_1$, $\alpha_2$, and $\alpha_3$ are positive dimensionless 
adjustable coefficients that depend on $(L, K_0)$.  It should be 
noticed that the dependence of $\Delta R/R$ on the EOS parameters is 
sufficiently weak that observations of $R/M$ would lead to deduction
of $\Delta R/R$ within $\sim 10 \%$ accuracy.

\begin{figure*}
\begin{center}
\begin{tabular}{ccc}
\includegraphics[scale=0.4]{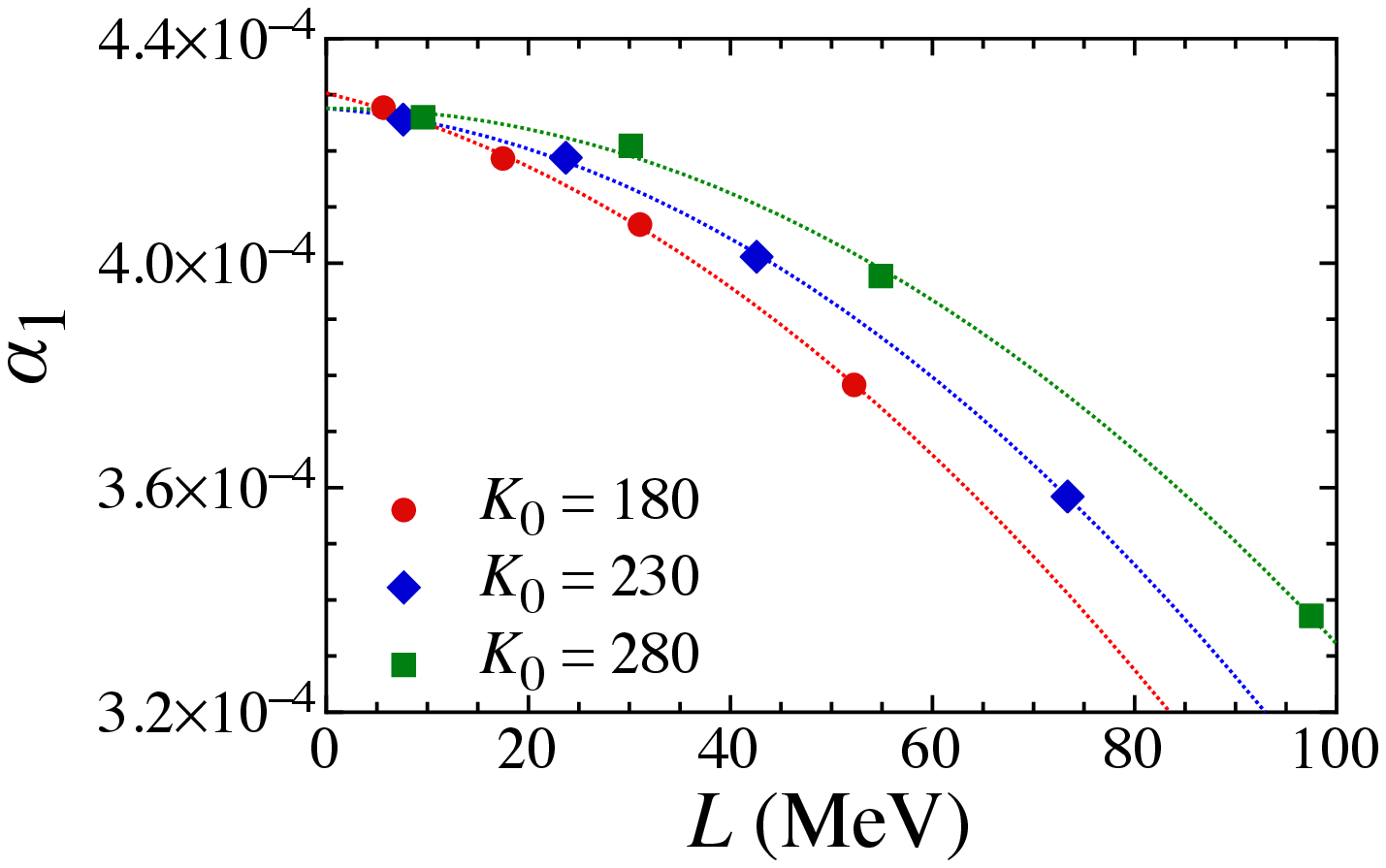} &
\includegraphics[scale=0.4]{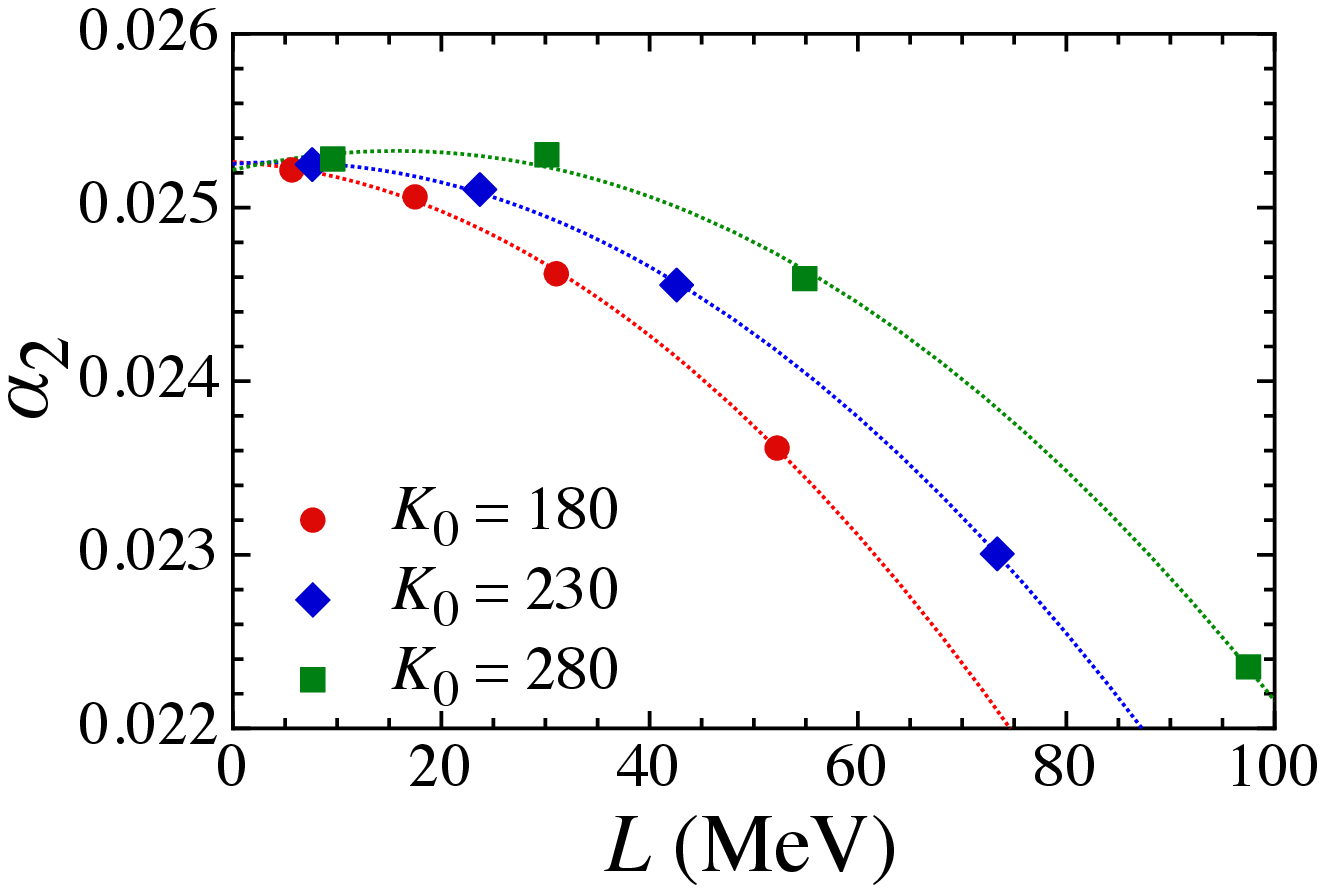} &
\includegraphics[scale=0.4]{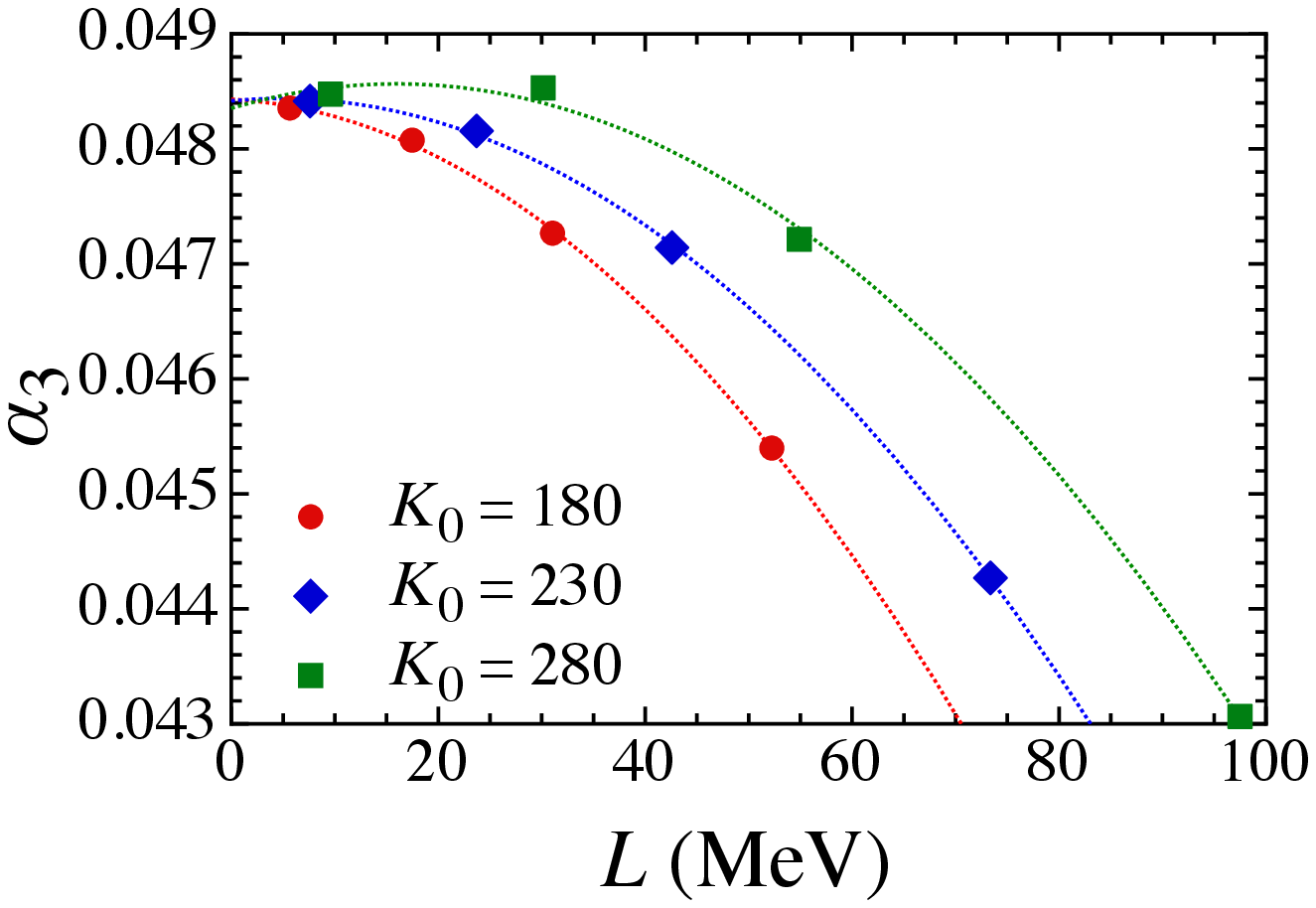} 
\end{tabular}
\end{center}
\caption{
Same as Fig.\ \ref{fig:asp-L}, but for the coefficients in Eq.\ 
(\ref{eq:dR-MR}) and the fitting formula given by Eqs.\ 
(\ref{eq:dR-a1})--(\ref{eq:dR-a3}).
}
\label{fig:a-L}
\end{figure*}

The coefficients in Eq.\ (\ref{eq:dR-MR}) obtained by fitting for various
sets of the EOS parameters are shown in Fig.\ \ref{fig:a-L} as a function 
of $L$.  Then, we can successfully derive the fitting formula as
\begin{eqnarray}
  \alpha_1 &=& \beta_{11} + \beta_{12}\left(\frac{L}{60\ {\rm MeV}}\right) + \beta_{13}\left(\frac{L}{60\ {\rm MeV}}\right)^2,
       \label{eq:dR-a1} \\
  \alpha_2 &=& \beta_{21} + \beta_{22}\left(\frac{L}{60\ {\rm MeV}}\right) + \beta_{23}\left(\frac{L}{60\ {\rm MeV}}\right)^2,
       \label{eq:dR-a2} \\  
  \alpha_3 &=& \beta_{31} + \beta_{32}\left(\frac{L}{60\ {\rm MeV}}\right) + \beta_{33}\left(\frac{L}{60\ {\rm MeV}}\right)^2, 
       \label{eq:dR-a3}  
\end{eqnarray}
where $\beta_{ij}$ with $i=1$, 2, 3 and $j=1$, 2, 3 are dimensionless 
adjustable coefficients that depend on $K_0$.  In deriving these 
fitting formulas, we omit the results with $K_0=360$ MeV, which are 
difficult to fit as in the cases of $\Delta R_{\rm sp}/R$ and 
$\Delta R_{\rm cy}/R$.

\begin{figure}
\begin{center}
\includegraphics[scale=0.5]{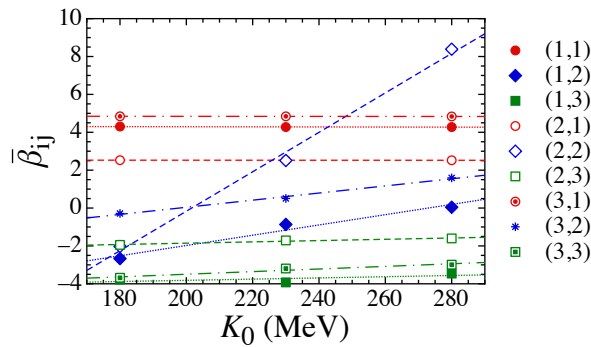}
\end{center}
\caption{
The normalized quantities $\bar{\beta}_{ij}$ of the coefficients in 
Eqs.\ (\ref{eq:dR-a1})--(\ref{eq:dR-a3}) plotted as a function of 
$K_0$, where the labels of $(i,j)$ for $i=1$, 2, 3 and $j=1$, 2, 3 denote 
the subscript in $\bar{\beta}_{ij}$.  The dotted lines denote the fitting formulas 
given by Eqs.\ (\ref{eq:dR-b11})--(\ref{eq:dR-b33}).
}
\label{fig:b-K0}
\end{figure}

Finally, in Fig.\ \ref{fig:b-K0} we exhibit the normalized quantities
$\bar{\beta}_{ij}$ of the coefficients in Eqs.\ 
(\ref{eq:dR-a1})--(\ref{eq:dR-a3}) as a function of $K_0$, where 
$\bar{\beta}_{ij}$ is given by 
$\bar{\beta}_{11}=\beta_{11}\times 10^4$, 
$\bar{\beta}_{12}=\beta_{12}\times 10^5$, 
$\bar{\beta}_{13}=\beta_{13}\times 10^5$, 
$\bar{\beta}_{21}=\beta_{21}\times 10^2$, 
$\bar{\beta}_{22}=\beta_{22}\times 10^4$, 
$\bar{\beta}_{23}=\beta_{23}\times 10^3$, 
$\bar{\beta}_{31}=\beta_{31}\times 10^2$, 
$\bar{\beta}_{32}=\beta_{32}\times 10^3$, 
and $\bar{\beta}_{33}=\beta_{33}\times 10^3$.  Again, the coefficients
$\beta_{ij}$ can be expressed as a linear function of $K_0$ by
\begin{eqnarray}
  \beta_{11} &=&   \left[  4.3474   -    0.06256\left(\frac{K_0}{230\ {\rm MeV}}\right)\right]\times 10^{-4},  \label{eq:dR-b11} \\
  \beta_{12} &=&   \left[ -7.4004   +   6.2366\left(\frac{K_0}{230\ {\rm MeV}}\right)\right]\times 10^{-5},   \label{eq:dR-b12} \\
  \beta_{13} &=&   \left[ -4.4550   +   0.7337\left(\frac{K_0}{230\ {\rm MeV}}\right)\right]\times 10^{-5},  \label{eq:dR-b13} \\
  \beta_{21} &=&   \left[  2.5350   -   0.01058\left(\frac{K_0}{230\ {\rm MeV}}\right)\right]\times 10^{-2},  \label{eq:dR-b21} \\
  \beta_{22} &=&   \left[ -20.9930  +   23.9487  \left(\frac{K_0}{230\ {\rm MeV}}\right)\right]\times 10^{-4},   \label{eq:dR-b22} \\
  \beta_{23} &=&   \left[ -2.5407    +  0.7880\left(\frac{K_0}{230\ {\rm MeV}}\right)\right]\times 10^{-3},  \label{eq:dR-b23} \\
  \beta_{31} &=&   \left[ 4.8581   - 0.01794 \left(\frac{K_0}{230\ {\rm MeV}}\right)\right]\times 10^{-2},  \label{eq:dR-b31} \\
  \beta_{32} &=&   \left[ -3.7203   +   4.3243\left(\frac{K_0}{230\ {\rm MeV}}\right)\right]\times 10^{-3},  \label{eq:dR-b32} \\
  \beta_{33} &=&   \left[ -4.8761    +  1.5884\left(\frac{K_0}{230\ {\rm MeV}}\right)\right]\times 10^{-3}.  \label{eq:dR-b33} 
\end{eqnarray}
Now, we obtain a complete set of the fitting formulas 
(\ref{eq:dR-MR})--(\ref{eq:dR-b33}), which well reproduce the calculated 
values of $\Delta R/R$ for various combinations of $R/M$, $L$, and $K_0$.  
Note that applicability of these formulas is here again 
limited to the range of $180\lsim K_0\lsim 280$ MeV.

\section{Conclusion}
\label{sec:IV}

We have constructed the fitting formulas for the thickness of the whole 
crust and the layers of the respective pasta phases in a manner that is 
dependent on the neutron star compactness $M/R$ and the EOS parameters 
$L$ and $K_0$.  We find from the approximate form of the EOS [Eq. (\ref{eq:w})] and 
the thickness [Eq. (\ref{eq:mvt})] that the $L$ and $K_0$ dependence is much weaker 
than the $M/R$ for the whole crust and the SP phase, while being as strong 
for each of the pasta (C, S, CH, SH) phases.  
We remark that the known approximate dependence of the crust thickness 
on the compactness (e.g., \cite{PR1995}) is updated here
by including the term of order $(M/R)^2$ in Eq.\ (\ref{eq:dR-MR}).
The resultant fitting formulas 
would be useful in deducing the thickness of the whole crust and the layer of 
the SP phase from future accurate determinations of $M/R$ of X-ray bursters 
by LOFT and millisecond pulsars by NICER.

We thank K. Nakazato for useful discussion in the initial stage of the present study. This work was supported in part by Grants-in-Aid for Scientific Research on Innovative Areas through No.\ 15H00843 and No.\ 24105008 provided by MEXT, by Grant-in-Aid for Young Scientists (B) through No.\ 26800133 provided by JSPS, and by Grant-in-Aid for Scientific Research (C) through Grant No. 17K05458 provided by JSPS.



\end{document}